\documentclass[aps,prd,onecolumn,showpacs,superscriptaddress,preprintnumbers,floatfix,reprint,nofootinbib]{revtex4-1}
\usepackage{graphicx}
\usepackage{amsmath}
\usepackage{bm}
\usepackage{yhmath}
\usepackage[colorlinks,citecolor=blue,urlcolor=blue,linkcolor=blue]{hyperref}
\usepackage{subfigure}
\usepackage{color}
\usepackage{cases}
\usepackage{dcolumn,booktabs,bm}
\usepackage{slashed}
\usepackage{amsfonts,amssymb,stmaryrd,latexsym,amsmath}
\usepackage{textcomp}
\usepackage{multirow}
\usepackage{cancel}
\usepackage{diagbox}
\usepackage{array}

\renewcommand{\arraystretch}{1.8}
\allowdisplaybreaks

\begin{document}

\title{Spectrum of the fully-heavy tetraquark state $QQ\bar Q' \bar Q'$}
\author{Guang-Juan Wang}\email{wgj@pku.edu.cn}
 \affiliation{Center of High Energy Physics, Peking University, Beijing 100871, China}
\affiliation{School of Physics and State Key Laboratory of Nuclear
Physics and Technology, Peking University, Beijing 100871, China}

\author{Lu Meng}\email{lmeng@pku.edu.cn}
\affiliation{School of Physics and State Key Laboratory of Nuclear
Physics and Technology, Peking University, Beijing 100871, China}

\author{Shi-Lin Zhu}\email{zhusl@pku.edu.cn}
\affiliation{School of Physics and State Key Laboratory of Nuclear
Physics and Technology, Peking University, Beijing 100871,
China}\affiliation{Collaborative Innovation Center of Quantum
Matter, Beijing 100871, China}

\begin{abstract}
In this work, we systematically calculate the mass spectra of the
$S$-wave fully heavy tetraquark states $bb\bar b\bar b$, $cc\bar
c\bar c$, and $bb\bar c\bar c$ in two nonrelativistic quark models.
A tetraquark state may be an admixture of a $6_c-\bar 6_c$ state and
a $\bar 3_c-3_c$ one, where $6_c-\bar 6_c$($\bar 3_c-3_c$) denotes
the color configuration with a $6_c$ ($\bar 3_c$) diquark and a
$\bar 6_c$ ($3_c$) antidiquark. For the tetraquark states $bb\bar
b\bar b$ and $cc\bar c\bar c$ with $J^{PC}={0^{++}}$, the $6_c-\bar
6_c$ state is lower than the $\bar 3_c-3_c$ one in both the two
quark models, while the order of the $bb\bar c\bar c$ states depend
on models. The $6_c-\bar 6_c$ and $\bar 3_c-3_c$ mixing effects are
induced by the hyperfine interactions between the diquark and
antidiquark, while the contributions from the one-gluon-exchange
(OGE) Coulomb or the linear confinement potentials vanish for the
$QQ\bar Q'\bar Q'$ system. With the couple-channel effects, we
obtain the similar mass spectra. The numerical results show that the
ground $QQ\bar Q'\bar Q'$ ($Q=b,c$ and $Q'=b,c$) tetraquark states
are located above the corresponding scattering states, which
indicates that there may not exist a bound state in the scheme of
the two quark models.

\end{abstract}
\pacs{~12.40.Yx,~14.40.Pq,~12.39.x} \maketitle

\section{introduction}\label{intro}
Since 2003, numerous exotic structures have been observed in
experiments~\cite{Choi:2007wga,
Aaij:2014jqa,Chilikin:2013tch,Chilikin:2014bkk,Ablikim:2013xfr,Ablikim:2013wzq,Ablikim:2013mio,Belle:2011aa,Adachi:2012cx,Aaij:2015tga,Aaij:2018bla},
amongst which many states cannot be accommodated into the
traditional quark model. In the literature, there are many possible
explanations. The most prominent ones are the molecules (loosely
bound states of two hadrons), the tetraquarks (compact bound
states), and the hybrids (composed of gluons and quarks), etc. For a
recent review, see
Refs.~\cite{Chen:2016qju,Guo:2017jvc,Esposito:2016noz,Ali:2017jda,Liu:2019zoy}.

A fully heavy tetraquark state is a topic of great interest. The
interactions between the heavy quarks may be dominated by the
short-range one-gluon-exchange (OGE) potential rather than the
long-range potentials. Thus, they are good candidates of the compact
tetraquark states. Unlike a meson or a baryon where the color
configuration of the quarks is unique, i.e. $q_i\bar q_j\delta_{ij}$
or $\epsilon_{ijk}q_iq_jq_k$, the color structure for the tetraquark
is much richer. For the tetraquark states, the four quarks can
neutralize the color in two ways, $6_c\otimes \bar 6_c=1_c$ and
$\bar 3_c\otimes 3_c=1_c$. In this work, we label the two color
configurations $|(QQ)_{6_c}{\bar Q\bar Q}_{\bar 6_c}\rangle$ and
$|(QQ)_{\bar 3_c}{\bar Q\bar Q}_{3_c}\rangle$ as $6_c-\bar 6_c$ and
$\bar 3_c-3_c$, respectively. In
Refs.~\cite{Maiani:2004vq,Ali:2011ug,Eichten:2017ffp}, the authors
investigate the tetraquark states in the $\bar 3_c-3_c$
configuration. In Refs.~\cite{Chen:2016oma,Maiani:2019cwl}, the
authors pointed out that the $6_c-\bar 6_c$ configuration is also
very important to form the tetraquark states. The fully heavy
tetraquark state is a golden system to investigate the inner color
configuration of the multiquark states. For the above reasons, the
fully heavy tetraquark states have inspired both the experimental
and theoretical attention.

Recently, the CMS collaboration observed the $\Upsilon(1S)$ pair
production and indicated a $bb\bar b\bar b$ signal around $18.4$ GeV
with a global significance of $ 3.6
\sigma$~\cite{Khachatryan:2016ydm,S. Durgut}. Later, the LHCb
searched the invariant mass distribution of $\Upsilon(1s)\mu^+\mu^-$
and did not observe the tetraquark state $X_{b b\bar b\bar
b}$~\cite{Aaij:2018zrb}. The tension between CMS and LHCb data
requires more experimental and theoretical studies of the
fully-beauty tetraquarks.

The mass spectroscopy has been a major platform to probe the
dynamics of the tetraquarks. Since 1975, there have been many
theoretical works about the mass spectroscopy of the fully heavy
quark
states~\cite{Iwasaki:1975pv,Chao:1980dv,Ader:1981db,Zouzou:1986qh,Heller:1986bt,SilvestreBrac:1992mv,SilvestreBrac:1993ry}.
The existence of the fully heavy quark states is still
controversial. Recent interests have followed the experimental
developments in the past several years. The mass spectra have been
calculated in different schemes, for instance, a diffusion Monte
Carlo method~\cite{Bai:2016int}, the non-relativistic effective
field theory (NREFT)~\cite{Anwar:2017toa}, the QCD sum
rules~\cite{Wang:2017jtz,Wang:2018poa,Chen:2018cqz}, covariant
Bethe-Salpeter equations~\cite{Heupel:2012ua}, various quark
models~\cite{Lloyd:2003yc,Debastiani:2017msn, Barnea:2006sd}, and
other phenomenological
models~\cite{Berezhnoy:2011xy,Berezhnoy:2011xn,Karliner:2016zzc,Esposito:2018cwh,Karliner:2017qhf}.
The lowest $bb\bar b\bar b$ and $cc\bar c\bar c$ states are
estimated to be in the mass range $18-20$ GeV and $5-7$ GeV,
respectively. In contrast, the authors of Ref.~\cite{Wu:2016vtq}
investigated the mass spectra of the $QQ\bar Q\bar Q$ states in the
Chromomagnetic interaction (CMI) model and concluded that no stable
$QQ\bar Q\bar Q$ states exist. Later, several other approaches, such
as the nonrelativistic chiral quark model~\cite{Chen:2019dvd,
Liu:2019zuc}, the lattice QCD~\cite{Hughes:2017xie} and other
models~\cite{Richard:2017vry,Czarnecki:2017vco} also do not support
the existence of the bound $QQ\bar Q\bar Q$ states.

To investigate the existence of the full heavy tetraquark states, we
systematically calculate the mass spectra of the $bb\bar b\bar b$,
$cc\bar c\bar c$ and $bb\bar c\bar c~(cc\bar b\bar b)$ in two
non-relativistic quark models. In general, a tetraquark state should
be an admixture of the two color configurations, $6_c-\bar 6_c$ and
$\bar 3_c-3_c$. In this work, with the couple-channel effects, we
perform the dynamical calculation of the mass spectra of the
tetraquark states and investigate the inner structures of the ground
states.

We organize the paper as follows. In Sec.~\ref{sec1}, we introduce
the formalism to calculate their mass spectra, including two
non-relativistic quark models, the construction of the wave
functions, and the analytical expressions of the Hamiltonian matrix
elements. In Sec.~\ref{sec2}, we present the numerical results and
discuss the couple-channel effects between the $\bar 3_c-3_c$ and
$6_c-\bar 6_c$ configurations. In Sec. \ref{sec3}, we compare our
results with those in other models and give a brief summary.

 \section{formalism}\label{sec1}
\subsection{Hamiltonian}
The nonrelativistic Hamiltonian of a $Q_1Q_2\bar Q_3 \bar Q_4$
tetraquark state reads
\begin{eqnarray}
H & =& \sum_{i=1}^{4}\frac{p_{j}^{2}}{2m_{j}}+\sum_{i<j}V_{ij}+\sum_{i}m_{i}\nonumber\\
 & =&\frac{p^{2}}{2u}+V_{I}+h_{12}+h_{34}
\end{eqnarray}
with
\begin{eqnarray}
&&V_{I}=V_{13}+V_{14}+V_{23}+V_{24},\label{vi}\\
&&h_{ij} = \frac{p_{ij}^{2}}{2u_{ij}}+V_{ij}+m_{i}+m_{j},\\
&&\mathbf{p}_{ij}=\frac{m_{i}\mathbf{p}_{j}-m_{j}\mathbf{p}_{i}}{m_{i}+m_{j}}, \,\,\ u_{ij}=\frac{m_i m_j}{m_i+m_j}, \\
&& m_{ij}=m_i+m_j, \,\,\ u={{m_{12}m_{34}}\over{m_{12}+m_{34}}}, \\
&& \mathbf{P}_{ij}=\mathbf{p}_i+\mathbf{p}_j, \,\,
\mathbf{p}=\frac{m_{12}\mathbf{P}_{34}-m_{13}\mathbf{P}_{24}}{m_{12}+m_{34}}.
\end{eqnarray}
where $\mathbf{p}_i$ and $m_i$ are the momentum and mass of the
$i$th quark. The kinematic energy of the center-of-mass system has
been excluded by the constraint $\sum^4_{i=1}\mathbf{p}_i=0$.
$V_{ij}$ is the potential between the $i$th and $j$th quarks. The
$u_{ij}$, $m_{ij}$, $\mathbf{p}_{ij}$, and $ \mathbf{P}_{ij}$ are
the reduced mass, total mass, relative momentum, and total momentum
of the $(ij)$ pair of quarks, respectively. The $u$ and $\mathbf{p}$
are the reduced mass and relative momentum between the $(12)$ and
$(34)$ quark pairs. $h_{12}$, $h_{34}$ and $V_I$ represent the
$(12)$ quark pair inner interaction, $(34)$ quark pair interaction
and interaction between the two pairs.

Since the heavy quark mass is large, the relativistic effect is less
important. We use a nonrelativistic quark model to describe the
interaction between two heavy quarks. The quark model proposed in
Ref.~\cite{Wong:2001td} contains one gluon exchange (OGE) plus a
phenomenological linear confinement interaction and the $V_{ij}$
reads
\begin{eqnarray}\label{qm}
&&V_{ij}(r_{ij})=\frac{\mathbf{\lambda}_{i}}{2}\frac{\mathbf{\lambda}_{j}}{2}\left(V_{\text{coul}}+V_{\text{conf}}+V_{\text{hyp}}+V_{\text{cons}}\right)\nonumber\\
&&=\frac{\lambda_{i}}{2}\frac{\lambda_{j}}{2}\left(\frac{\alpha_{s}}{r_{ij}}-\frac{3b}{4}r_{ij}-\frac{8\pi\alpha_{s}}{3m_{i}m_{j}}\mathbf{s}_{i}\cdot\mathbf{s}_{j}e^{-\tau^{2}r^{2}}\frac{\tau^{3}}{\pi^{3/2}}+V_{\text{cons}}\right),\nonumber\\
\end{eqnarray}
where $\lambda$ is the color matrix (replaced by $-\lambda^{*}$ for
an antiquark). $\mathbf{s}_{i}$ is the spin operator of the $i$th
quark. $r_{ij}$ is the relative position of the $i$th and $j$th
quarks. $V_{\text{coul}}$, $V_{\text{conf}}$, and $V_{\text{hyp}}$
represent the OGE color Coulomb, the linear confinement, and the
hyperfine interactions, respectively. The OGE interaction leads to a
contact hyperfine effect and an infinite hyperfine splitting. In
Eq.~(\ref{qm}), the smearing effect has been considered in
$V_{\text{hyp}}$.

The $\alpha_{s}$ is the running coupling constant in the
perturbative QCD,
\begin{eqnarray}
\alpha_{s}(Q^{2})&=&\frac{12\pi}{(33-2N_{f})\ln(A+Q^{2}/B^{2})}.
\end{eqnarray}
In this work, we take the square of the invariant mass of the
interacting quarks as the scale $Q^{2}$. The values of the
parameters are listed in Table~\ref{par}. They are determined by
fitting the mass spectra of the mesons as listed in Table~\ref{meson
mass}.

 \begin{table*}
 \renewcommand\arraystretch{1.5}
 \caption{The values of parameters in quark model I~\cite{Wong:2001td} and model II~\cite{SilvestreBrac:1996bg}. }\label{par}
 \centering
 \setlength{\tabcolsep}{2.3mm}
\begin{tabular}{c|cccccccccc}
\toprule[1pt]\toprule[1pt] \multirow{2}{*}{Model I} &
\multirow{2}{*}{} & \multirow{2}{*}{} & $m_{c}${[}GeV{]} &
$m_{b}${[}GeV{]} & $b[\text{GeV}^{2}]$ & $\tau${[}GeV{]} &
$V_{\text{cons}}${[}GeV{]} & $A$ & $B${[}GeV{]} & \tabularnewline
 & & & $1.776$ & $5.102$ & $0.18$ & $0.897$ & $0.62$ & $10$ & $0.31$ & \tabularnewline
\midrule[1pt] \multirow{2}{*}{Model II} & $p$ & $r_{c}$ &
$m_{c}${[}GeV{]} & $m_{b}${[}GeV{]} & $\kappa$ & $\kappa'$ &
$\lambda[\text{GeV}^2]$ & $\Lambda${[}GeV{]} & $A[\text{GeV}^{B-1}]$
& $B$\tabularnewline
 & $1$ & $0$ & $1.836$ & $5\text{.}227$ & $0.5069$ & $1.8609$ & $0.1653$ & $0.8321$ & $1.6553$ & $0.2204$\tabularnewline
\bottomrule[1pt]\bottomrule[1pt]
\end{tabular}
\end{table*}

 \begin{table*}
 \renewcommand\arraystretch{1.5}
\caption{The mass spectra of the heavy quarkonia in units of MeV.
The $M_{\text{ex}}$, $M^I_{th}$, and $M^{II}_{th}$ refer to the mass
spectra of mesons from experiments~\cite{Tanabashi:2018oca}, in
model I~\cite{Wong:2001td}, and in model
II~\cite{SilvestreBrac:1996bg}, respectively. }\label{meson mass}
 \centering
 \setlength{\tabcolsep}{3mm}
\begin{tabular}{cccc|ccccc}
\toprule[1pt]\toprule[1pt]
 & $M_{\text{ex}}$ & $M_{th}^{I} $& $M_{th}^{II}$ & & $M_{\text{ex}}$& $M_{th}^{I}$ & $M_{th}^{II}$& \tabularnewline
\midrule[1pt]
 $B_{c}$ & $6274.9$ & $6319.4$ & $6293.5$ & & & & \tabularnewline
 $\eta_{c}$ & $2983.9$ & $3056.5$ & $3006.6$ & $\eta_{b}$ & $9399.0$ & $9497.8$ & $9427.9$\tabularnewline

$\eta_{c}(2S)$ & $3637.6$ & $3637.6$ & $3621.2$ & $\Upsilon(1S)$ &
$9460.30$ & $9503.6$ & $9470.4$\tabularnewline

$J/\psi$ & $3096.9$ & $3085.1$ & $3102.1$ & $\Upsilon(2S)$ &
$10023.26$ & $9949.7$ & $10017.8$\tabularnewline

$\psi(2S)$ & $3686.1$ & $3652.4$ & $3657.8$ & $\Upsilon(3S)$ &
$10355.2$ & $10389.8$ & $10440.6$\tabularnewline
\bottomrule[1pt]\bottomrule[1pt]
\end{tabular}
\end{table*}

To investigate the model dependence of the mass spectrum, we also
consider another nonrelativistic quark model proposed in
Ref.~\cite{SilvestreBrac:1996bg}. The potential reads
\begin{eqnarray}
&&V_{ij}(r_{ij})=-\frac{3}{16}\sum_{i<j}\lambda_{i}\lambda_{j}\Big(-\frac{\kappa(1-\text{exp}(-r_{ij}/r_{c}))}{r_{ij}}+\lambda r_{ij}^{p}\nonumber \\
&&-\Lambda+\frac{8\pi}{3m_{i}m_{j}}\kappa'(1-\text{exp}(-r_{ij}/r_{c}))\frac{\text{exp}(-r_{ij}^{2}/r_{0}^{2})}{\pi^{3/2}r_{0}^{3}}\mathbf{s}_{i}\cdot\mathbf{s}_{j}\Big),\nonumber \\
\end{eqnarray}
where $r_0=A(\frac{2m_im_j}{m_i+m_j})^{-B}$ is related to the
reduced mass of the two quarks $(ij)$. In this model, all the mass
information is included in the hyperfine potential, which is
expected to play a more important role than that in Model I. The
parameters of the potentials are listed in Table~\ref{par}. With
these parameters, we calculate the mass spectra of the mesons and
list them in Table~\ref{meson mass}.

In this work, we concentrate on the $S$-wave tetraquark states and
do not include the tensor and spin-orbital interactions in the two
quark models. In Table~\ref{meson mass}, we notice that both models
are able to reproduce the mass spectra of the heavy quarkonia. In
the following, we will extend the two quark models to study the
fully heavy tetraquarks.

\subsection{Wave function}\label{secwavefun}

\begin{figure*}[htbp]
\centering
\includegraphics[width=0.8\textwidth]{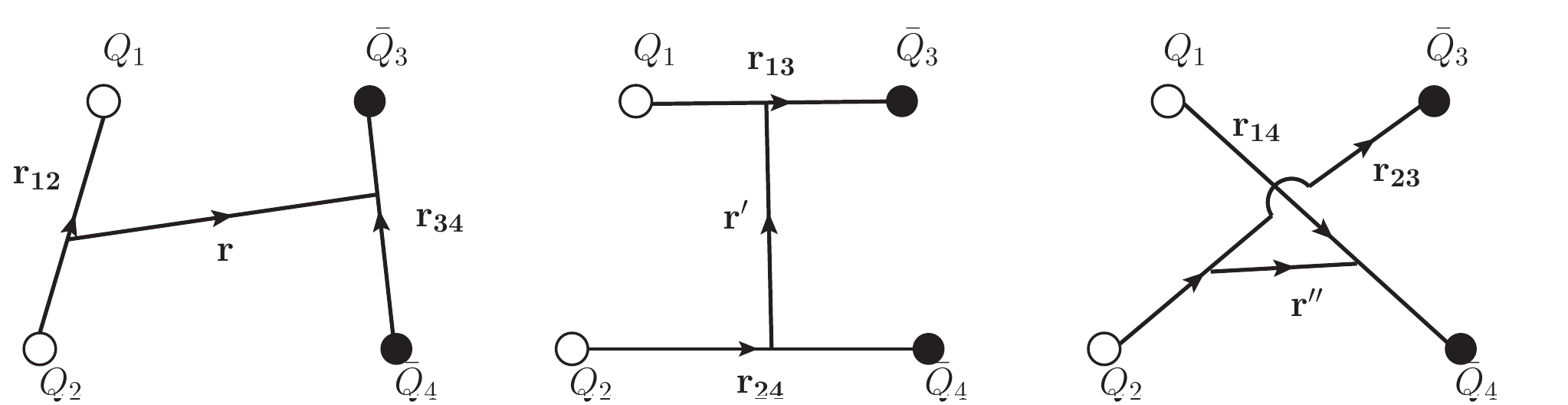}
\caption{The Jacobi coordinates in the tetraquark state. }
\label{jac}
\end{figure*}
In a $Q_1Q_2\bar Q_3\bar Q_4$ tetraquark state, there are three sets
of Jacobi coordinates as illustrated in Fig.~\ref{jac}. Each of them
contains three independent Jacobi coordinates, and they can be
transformed into others as follows,
\begin{eqnarray}\label{cor}
&&\mathbf{r}_{jk}=\mathbf{r}_j-\mathbf{r}_k=\mathbf{r}+c_{jk}^{a}\mathbf{r}_{12}+c_{jk}^{b}\mathbf{r}_{34},\nonumber\\
&&\mathbf{r}=\frac{ m_{1}\mathbf{r}_{1}+m_{2}\mathbf{r}_{2}}{m_{1}+m_{2}}-\frac{m_{3}\mathbf{r}_{3}+m_{4}\mathbf{r}_{4}}{m_{3}+m_{4}},\nonumber\\
&&\mathbf{r}'=\frac{m_{1}\mathbf{r}_{1}+m_{3}\mathbf{r}_{3}}{m_{1}+m_{3}}-\frac{m_{2}\mathbf{r}_{2}+m_{4}\mathbf{r}_{4}}{m_{2}+m_{4}}\nonumber\\
&&=\frac{(m_{1}m_{3}-m_{2}m_{4})\mathbf{r}+M_{T}u_{12}\mathbf{r}_{12}-M_{T}u_{34}\mathbf{r}_{34}}{(m_{1}+m_{4})(m_{2}+m_{3})},\nonumber\\
&&\mathbf{r}''=\frac{m_{1}\mathbf{r}_{1}+m_{4}\mathbf{r}_{4}}{m_{1}+m_{4}}-\frac{m_{2}\mathbf{r}_{2}+m_{3}\mathbf{r}_{3}}{m_{2}+m_{3}}\nonumber\\
&&=\frac{({m_1} {m_4}-m_2 m_3)\mathbf{r}+M_Tu_{12}\mathbf{r}_{12}
-M_T u_{34}\mathbf{r}_{34}}{(m_1+m_3) (m_2+m_4)},
\end{eqnarray}
where $M_T=\sum^4_{i=1}m_i$ is the total mass of the four quarks.
The transformation coefficients $c^{a(b)}_{jk}$ are listed in
Table~\ref{coe}. The superscripts $a$ and $b$ represent the quark
cluster and antiquark cluster, respectively.
\begin{table*}
\renewcommand\arraystretch{2.0}
\caption{The coefficient $c_{ij}$ in Eq. (\ref{cor}).}\label{coe}
 \setlength{\tabcolsep}{2.5mm}
\begin{tabular}{cccccccc}
\toprule[1pt]\toprule[1pt] $c_{14}^{a}$ & $c_{13}^{a}$ &
$c_{23}^{a}$ & $c_{24}^{a}$ & $c_{14}^{b}$ & $c_{13}^{b}$ &
$c_{23}^{b}$ & $c_{24}^{b}$\tabularnewline
 \midrule[1pt]
$\frac{m_{2}}{m_{1}+m_{2}}$ & $\frac{m_{2}}{m_{1}+m_{2}}$ &
$-\frac{m_{1}}{m_{1}+m_{2}}$ & $-\frac{m_{1}}{m_{1}+m_{2}}$ &
$\frac{m_{3}}{m_{3}+m_{4}}$ & $-\frac{m_{4}}{m_{3}+m_{4}}$ &
$-\frac{m_{4}}{m_{3}+m_{4}}$ &
$\frac{m_{3}}{m_{3}+m_{4}}$\tabularnewline
\bottomrule[1pt]\bottomrule[1pt]
\end{tabular}
\end{table*}

To simplify the calculation, we use the first coordinate
configuration to construct the wave function considering the
symmetry of the inner quarks. The wave function of a tetraquark
state is
\begin{eqnarray}\label{wavefunction}
&&\psi_{JJ_z}=\nonumber\\
&&\sum\left[\varphi_{n_a J_{a}}(\mathbf{r}_{12},\beta_a)\otimes\varphi_{n_b J_{b}}(\mathbf{r}_{34},\beta_b)\otimes \phi_{NL_{ab}}(\mathbf{r},\beta)\right]_{JJ_{z}},\nonumber\\
&&\varphi_{n_aJ_{a}M_{a}}=\left[\phi_{n_al_a}(\mathbf{r}_{12},\beta_a)\chi_{s_{a}}\right]_{M_{a}}^{J_{a}}\chi_{f}\chi_{c_a},
\end{eqnarray}
where the $\psi$ is the total wave function of the tetraquark state,
and $\varphi$ denotes that of the cluster (a) or (b). $J$ ($J_z$) is
the total angular momentum (the third direction component) of a
tetraquark state. The $\sum$ is the sum over all the possible wave
functions which may couple to the definite angular momentum $J$.
$n_{a(b)}$ and $N$ specify the radial dependence. The $s_{a(b)}$,
$l_{a(b)}$ and $J_{a(b)}$ are the spin, orbital and total angular
momentum of the cluster $a$ ($b$). $L_{ab}$ is the orbital angular
momentum between the two clusters. The $\chi_{s}$, $\chi_f$,
$\chi_c$ are the wave functions in the spin, the isospin, and the
color space, respectively. $\phi$ is the spatial wave function and
is expressed by the Gaussian basis~\cite{Hiyama:2003cu},
\begin{eqnarray}
\phi_{n_a l_a{m_a}}(\mathbf{r}_{12},\beta_a)&=&i^{l_{a}}r_{12}^{l_{a}}\sqrt{\frac{4\pi}{(2l_{a}+1)!!}}(\frac{n_a\beta_{a}^{2}}{\pi})^{3/4}\nonumber\\
&\times&(2n_a\beta_{a}^{2})^{l_{a}/2}e^{-r^{2}\beta_{a}^{2}n_a/2}Y_{l_{a}m_{a}}(\mathbf{\Omega}_{12}).\nonumber
\end{eqnarray}
with $\beta_{a}$ being the oscillating parameter.

In this work, we concentrate on the $S$-wave tetraquark states.
Their wave functions are expanded by the basis which satisfies the
relation $\mathbf{l}_a+\mathbf{l}_b+\mathbf{L}_{ab}=\mathbf{0}$. The
states with higher orbital excitations contribute to the ground
state through the tensor or the spin-orbital potentials. These
contributions are higher order effects and neglected in this work.
Thus, for the lowest $S$-wave tetraquark states, we only consider
the wave functions with $l_a=l_b={L}_{ab}=0$. The wave function of
the tetraquark state in Eq. (\ref{wavefunction}) is simplified as
\begin{eqnarray}\label{wf}
\psi_{SS_z}&=&\sum_{\alpha_,n_A,n_b,n_{ab}} \chi_{\alpha} \phi_{n_a}(\mathbf{r}_{12},\beta_a)\phi_{n_b}(\mathbf{r}_{34},\beta_b) \phi_{n_{ab}}(\mathbf{r},\beta),\nonumber\\
 \chi_{\alpha} &=& \left[\chi_{s_a}\otimes \chi_{s_b}\right]^S\left[\chi_{f_a}\otimes \chi_{f_b}\right]\left[\chi_{c_a}\otimes \chi_{c_b}\right]^1,
\end{eqnarray}
where $S$ is the total spin of the tetraquark state and $1$
represents the color-singlet representation. For the spatial wave
functions, we have omitted the orbital angular momentum in the
Gaussian wave function $\phi$.

The wave functions are constrained by the Pauli principle. The
$S$-wave diquark (antidiquark) with two identical quarks
(antiquarks) has two possible configurations as listed in
Table~\ref{con}. Then, for the $cc\bar c \bar c$, $bb\bar b\bar b$,
and $bb \bar c \bar c$ tetraquark states, the possible
color-flavor-spin functions read
\begin{itemize}
    \item   $J^{PC}=0^{++}$
    \begin{eqnarray}\label{0p}
    & \chi_{1}=\left[[QQ]_{\bar{3}_{c}}^{1}[\bar{Q}\bar{Q}]_{3_{c}}^{1}\right]_{1_{c}}^{0},\,\,\, \chi_{2}=\left[[QQ]_{6_{c}}^{0}[\bar{Q}\bar{Q}]_{\bar{6}_{c}}^{0}\right]_{1_{c}}^{0}.
    \end{eqnarray}

\item   $J^{PC}=1^{+-}$
    \begin{eqnarray}\label{1p}
    \chi_{1}=\left[[QQ]_{\bar{3}_{c}}^{1}[\bar{Q}\bar{Q}]_{3_{c}}^{1}\right]_{1_{c}}^{1}.
    \end{eqnarray}

\item   $J^{PC}=2^{++}$
    \begin{eqnarray}
    \label{2p}
    \chi_{1}=\left[[QQ]_{\bar{3}_{c}}^{1}[\bar{Q}\bar{Q}]_{3_{c}}^{1}\right]_{1_{c}}^{2}.
    \end{eqnarray}
\end{itemize}
where the superscript and subscript denote the spin and color
representations.

\subsection{Hamiltonian matrix elements}

\begin{table}
 \renewcommand\arraystretch{1.5}
\caption{The configurations of the diquark (antiquark) constrained
by Pauli principle. ``S'' and ``A'' represent symmetry and
antisymmetry.}\label{con}
 \setlength{\tabcolsep}{2.5mm}
\begin{tabular}{lc|lccc}
\toprule[1pt] \toprule[1pt] $J^{P}=1^{+}$ & $QQ$ & $J^{P}=0^{+}$ &
$QQ$\tabularnewline \midrule[1pt] $S$-wave(L=0) & S & $S$-wave(L=0)
& S \tabularnewline

Flavor & S & Flavor & S\tabularnewline

Spin(S=1) & S & Spin(S=0) & A \tabularnewline

Color($\bar{3}_c$) & A & Color($6_c$) & S\tabularnewline
\bottomrule[1pt] \bottomrule[1pt]
\end{tabular}
\end{table}
With the wave function constructed in section~\ref{secwavefun}, we
calculate the Hamiltonian matrix elements. For the quark model I,
the matrix element of $\langle h_{12}\rangle$ reads,
 \begin{eqnarray}
&&\langle\chi_{\alpha_i}\phi_{n}(r_{12})\phi_{\lambda}(r_{34})\phi_{k}(r)|h_{12}|\chi_{\alpha_j}\phi_{m}(r_{12})\phi_{\nu}(r_{34})\phi_{k'}(r)\rangle\nonumber\\
&&=\delta_{\alpha_i\alpha_j}N_{\lambda,\nu}N_{k,k'}\langle\phi_{n}(r_{12},\beta_a)|h_{12}|\phi_{m}(r_{12},\beta_a)\rangle \nonumber\\
&&=\delta_{\alpha_i\alpha_j}N_{\lambda,\nu}N_{k,k'}\left(\langle
T_{12}+ m_1+m_2\rangle+ \langle V_{12}\rangle \right),
 \end{eqnarray}
with
 \begin{eqnarray}
&&N_{k,k'}=\left(\frac{2\sqrt{kk'}}{k+k'}\right)^{3/2},\nonumber\\
&&\langle T_{12}+ m_1+m_2\rangle=N_{m,n}\left(\frac{3mn\beta_a^{2}}{2u_{12}(m+n)}+m_1+m_2\right),\nonumber\\
&&\langle V_{12}(\mathbf{r}_{12})\rangle=\langle V_{\text{coul}}\rangle+\langle V_{\text{conf}}\rangle+\langle V_{\text{hyp}}\rangle+\langle V_{\text{cons}} \rangle,\nonumber\\
&&\langle V_{\text{coul}}\rangle=I_{C}\frac{4\pi\alpha_{s}\beta_a}{(2\pi)^{3/2}}\sqrt{m+n}N_{m,n},\nonumber\\
&&\langle V_{\text{conf}}\rangle=-\frac{3}{4}I_{C}\frac{8\pi b}{(2\pi)^{3/2}\beta_a\sqrt{m+n}}N_{m,n},\nonumber\\
&&\langle V_{\text{hyp}}\rangle=-I_{CM}\frac{8\pi\alpha_{s}}{3m_{i}m_{j}}\frac{\sigma^{3}}{\pi^{3/2}}\left(\frac{2\sqrt{mn}}{m+n+2\sigma^{2}/\beta_a^{2}}\right)^{\frac{3}{2}},\nonumber\\
&&\langle V_{\text{cons}} \rangle=I_{C}V_{\text{cons}}N_{m,n},
 \end{eqnarray}
 where $n,\lambda,k,m,\nu,k'$ specify the radial dependence. The $I_C$ and $I_{\text{CM}}$ are the color factor and the color electromagnetic factor in Table~\ref{color} and Table~\ref{Icm}, respectively. $\chi_{\alpha_i,\alpha_j}$ denote the color-flavor-spin configurations as illustrated in Eq. (\ref{wf}). Since the potential $h_{12}$ is diagonal in the color-flavor-spin space, it does not induce the coupling of different $\chi_{\alpha_i,\alpha_j}$ channels and the $\langle h_{12}\rangle $ is proportional to $\delta_{\alpha_i\alpha_j}$. The derivation of $\langle h_{34}\rangle$ is similar to that of $\langle h_{12}\rangle $.

 \begin{table}
 \renewcommand\arraystretch{1.5}
 \caption{The color matrix element $I_C=\langle\frac{\mathbf{\lambda}_{i}}{2}\frac{\mathbf{\lambda}_{j}}{2}\rangle$ for the $(ij)$ pair of quarks. The subscripts denote the color representation of the cluster.}\label{color}
 \setlength{\tabcolsep}{2.5mm}
 \begin{tabular}{cccccccc}
\toprule[1pt]\toprule[1pt]
\multicolumn{6}{c}{$\langle(Q_{1}Q_{2})_{\bar
3}(\bar{Q}_{3}\bar{Q}_{4})_3|\frac{\mathbf{\lambda}_{i}}{2}\frac{\mathbf{\lambda}_{j}}{2}|(Q_{1}Q_{2})_{\bar
3}(\bar{Q}_{3}\bar{Q}_{4})_3\rangle$} \tabularnewline \midrule[1pt]
 $Q_{1}\bar{Q}_{3}$ & $Q_{2}\bar{Q}_{4}$ & $Q_{1}\bar{Q}_{4}$ & $Q_{2}\bar{Q}_{3}$ & $Q_{1}Q_{2}$ & $\bar{Q}_{3}\bar{Q}_{4}$\tabularnewline
 $-\frac{1}{3}$ & $-\frac{1}{3}$ & $-\frac{1}{3}$ & $-\frac{1}{3}$ & $-\frac{2}{3}$ & $-\frac{2}{3}$\tabularnewline
\midrule[1pt]
\multicolumn{6}{c}{$\langle(Q_{1}Q_{2})_{6}(\bar{Q}_{3}\bar{Q}_{4})_{\bar
6}|\frac{\mathbf{\lambda}_{i}}{2}\frac{\mathbf{\lambda}_{j}}{2}|(Q_{1}Q_{2})_{6}(\bar{Q}_{3}\bar{Q}_{4})_{\bar
6}\rangle$} \tabularnewline \midrule[1pt]
 $Q_{1}\bar{Q}_{3}$ & $Q_{2}\bar{Q}_{4}$ & $Q_{1}\bar{Q}_{4}$ & $Q_{2}\bar{Q}_{3}$ & $Q_{1}Q_{2}$ & $\bar{Q}_{3}\bar{Q}_{4}$\tabularnewline

 $-\frac{5}{6}$ & $-\frac{5}{6}$ & $-\frac{5}{6}$ & $-\frac{5}{6}$ & $\frac{1}{3}$ & $\frac{1}{3}$\tabularnewline
 \midrule[1pt]
\multicolumn{6}{c}{$\langle(Q_{1}Q_{2})_{\bar
3}(\bar{Q}_{3}\bar{Q}_{4})_{3}|\frac{\mathbf{\lambda}_{i}}{2}\frac{\mathbf{\lambda}_{j}}{2}|(Q_{1}Q_{2})_{6}(\bar{Q}_{3}\bar{Q}_{4})_{\bar
6}\rangle$} \tabularnewline \midrule[1pt]

 $Q_{1}\bar{Q}_{3}$ & $Q_{2}\bar{Q}_{4}$ & $Q_{1}\bar{Q}_{4}$ & $Q_{2}\bar{Q}_{3}$ & $Q_{1}Q_{2}$ & $\bar{Q}_{3}\bar{Q}_{4}$\tabularnewline
 $-\frac{1}{\sqrt{2}}$ & $-\frac{1}{\sqrt{2}}$ & $\frac{1}{\sqrt{2}}$ & $\frac{1}{\sqrt{2}}$ & $0$ & $0$\tabularnewline
 \bottomrule[1pt]\bottomrule[1pt]
\end{tabular}
\end{table}

 \begin{table}
 \renewcommand\arraystretch{1.5}
 \caption{The color magnetic factor $\langle I_{\text{CM}}\rangle_{\alpha_i \alpha_j}=\langle \chi_{\alpha_i}|\frac{\lambda_{i}}{2}\frac{\lambda_{j}}{2}\mathbf{s}_{i}\cdot \mathbf{s}_{j}|\chi_{\alpha_j}\rangle$ for the $(ij)$ quark pairs. The $\chi_{\alpha_{i,j}}$ denotes the color-flavor-spin wave functions in Eqs. (\ref{0p})-(\ref{2p}).}\label{Icm}
 \setlength{\tabcolsep}{2.5mm}
\begin{tabular}{c|ccccc}
\toprule[1pt]\toprule[1pt]
 & \multicolumn{3}{c}{$I_{CM}^{ij}=\langle\frac{\lambda_{i}}{2}\frac{\lambda_{j}}{2}s_{i}\cdot s_{j}\rangle$}\tabularnewline
\midrule[1pt] \multirow{6}{*}{$0^{++}$} & $\langle
I_{CM}^{Q\bar{Q}}\rangle_{11}$ & $\langle I_{CM}^{QQ}\rangle_{11}$ &
$\langle I_{CM}^{\bar{Q}\bar{Q}}\rangle_{11}$\tabularnewline
 & $\frac{1}{6}$ & $-\frac{1}{6}$ & $-\frac{1}{6}$\tabularnewline
 & $\langle I_{CM}^{Q\bar{Q}}\rangle_{22}$ & $\langle I_{CM}^{QQ}\rangle_{22}$ & $\langle I_{CM}^{QQ}\rangle_{22}$\tabularnewline
 & $0$ & $-\frac{1}{4}$ & $-\frac{1}{4}$\tabularnewline
 & $\langle I_{CM}^{Q\bar{Q}}\rangle_{12}$ & $\langle I_{CM}^{QQ}\rangle_{11}$ & $\langle I_{CM}^{\bar{Q}\bar{Q}}\rangle_{11}$\tabularnewline
 & $\frac{\sqrt{3}}{4\sqrt{2}}$ & $0$ & $0$\tabularnewline
\midrule[1pt] \multirow{2}{*}{$1^{+-}$} & $\langle
I_{CM}^{Q\bar{Q}}\rangle_{11}$ & $\langle I_{CM}^{QQ}\rangle_{11}$ &
$\langle I_{CM}^{\bar{Q}\bar{Q}}\rangle_{11}$\tabularnewline
 & $\frac{1}{12}$ & $-\frac{1}{6}$ & $-\frac{1}{6}$\tabularnewline
\midrule[1pt] \multirow{2}{*}{$2^{++}$} & $\langle
I_{CM}^{Q\bar{Q}}\rangle_{11}$ & $\langle H_{CM}^{QQ}\rangle_{11}$ &
$\langle H_{CM}^{\bar{Q}\bar{Q}}\rangle_{11}$\tabularnewline
 & $-\frac{1}{12}$ & $-\frac{1}{6}$ & $-\frac{1}{6}$\tabularnewline
\bottomrule[1pt]\bottomrule[1pt]
\end{tabular}
\end{table}

Unlike the $h_{12}$ and $h_{34}$, the $V_I(\mathbf{r}_{ij})$ with
$i=1,2$ and $j=3,4$, which is the interaction between the diquark
and antidiquark, may lead to the mixing between different
color-spin-flavor configurations, i.e. $\chi_{\alpha_i}$ and
$\chi_{\alpha_j}$. The $\langle V_I(\mathbf{r}_{ij})\rangle$ reads
\begin{eqnarray*}
&&\langle\chi_{\alpha_i}\phi_{n}(r_{12},\beta_a)\phi_{\lambda}(r_{34},\beta_b)\phi_{k}(r,\beta)|V_I(\mathbf{r}_{ij})\nonumber \\
&&|\chi_{\alpha_j}\phi_{m}(r_{12},\gamma_a)\phi_{\nu}(r_{34},\gamma_b)\phi_{k'}(r,\gamma)\rangle \nonumber \\
&&=\langle
V_{\text{coul}}(\mathbf{r}_{ij})\rangle_{\alpha_i\alpha_j}+\langle
V_{\text{conf}}(\mathbf{r}_{ij})\rangle_{\alpha_i\alpha_j}+\langle
V_{\text{hyp}}(\mathbf{r}_{ij})\rangle_{\alpha_i\alpha_j},
\end{eqnarray*}
where ${\beta_{(a,b)}}$ and $\gamma_{{(a,b)}}$ are the oscillating
parameters. The implicit forms of the notations are

\begin{eqnarray}
&&\langle V_{\text{coul}}(\mathbf{r}_{ij})\rangle=I_C\tilde{N}_{m,n}\tilde{N}_{\lambda,\nu}\tilde{N}_{k,k'}\frac{2\alpha_{s}}{\sqrt{\pi}\sqrt{\frac{2}{k\beta^2+k'\gamma^2}+4a_{ij}^{2}}},\nonumber\\
&&\langle V_{\text{conf}}(\mathbf{r}_{ij})\rangle=I_C\tilde{N}_{m,n}\tilde{N}_{\lambda,\nu}\tilde{N}_{k,k'}(-\frac{3bz_{ij}}{\sqrt{\pi}}),\nonumber\\
&&\langle V_{\text{hyp}}(\mathbf{r}_{ij})\rangle =I_{CM}\tilde{N}_{m,n}\tilde{N}_{\lambda,\nu}{\tilde{N}_{k,k'}}\nonumber\\
&&~~~~~~~~~~~~~\times\left(-\frac{8\alpha_{s}}{3m_{i}m_{j}(4\tau_{ij}^{2}+\frac{2}{k\beta^2+k'\gamma^2})^{3/2}\sqrt{\pi}}\right),
\end{eqnarray}
where
\begin{eqnarray}
&&\tilde{N}_{m,n}=\left({2\sqrt{mn}\beta_a\gamma_a\over{m\gamma_a^2+n\beta_a^2}}\right)^{3/2},\\
&&a_{ij}=\frac{(c_{ij}^{a})^{2}}{2(m\beta_{a}^{2}+n{\gamma_a}^2)}+\frac{(c_{ij}^{b})^{2}}{2(\lambda\beta_{b}^{2}+\nu\gamma_{b}^{2})},\\
&& \tau_{ij}^{2}=a_{ij}^{2}+\frac{1}{4\sigma^{2}},\,\,
z_{ij}^{2}=a_{ij}^{2}+\frac{1}{2(k\beta^{2}+k'\gamma^{2})}.
\end{eqnarray}
With the above analytical expressions, we calculate the mass
spectrum of the fully heavy tetraquark states $QQ\bar Q'\bar Q'$.
The numerical results are given in the next section.

\section{Numerical results}\label{sec2}
\begin{figure}[htbp]
\centering
\includegraphics[width=.48\textwidth]{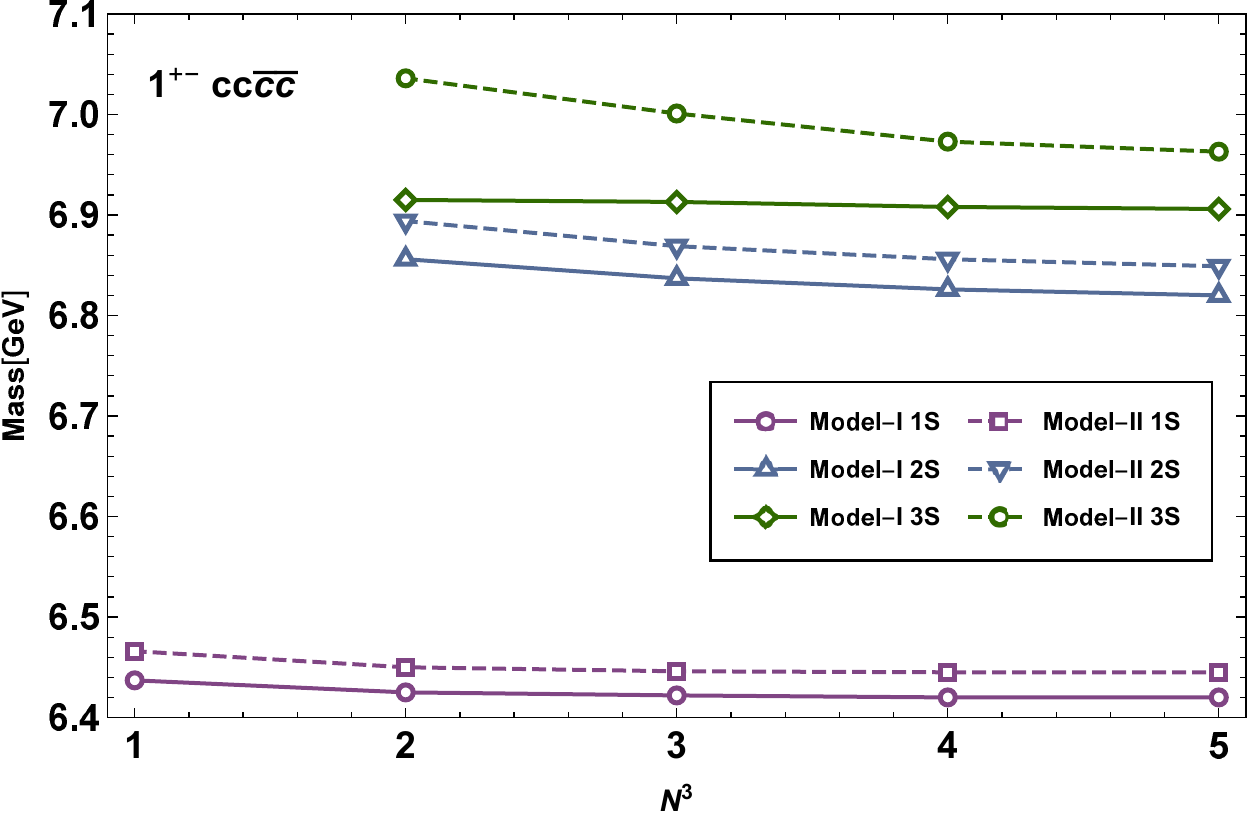}
\caption{The dependence of the mass spectrum on the number of
Gaussian basis $N^3$. The line and dashed line represent the
numerical results in model I and model II, respectively.}
\label{convergence}
\end{figure}

The wave function of a tetraquark state is composed of all wave
functions which subject to the conditions discussed in
section~\ref{secwavefun}. The number of the basis $N^3$ increases
from the minimum required to a large limit. We take the $cc\bar c
\bar c$ tetraquark state with $J^{PC}=1^{+-}$ as an example to
investigate the dependence of the results on the number of the
basis. Its wave function is expanded with $N^3=1^3$, $2^3$, $3^3$,
$4^3$ and $5^3$ basis, respectively. The corresponding eigenvalues
obtained through the variational method are displayed in
Fig.~\ref{convergence}. The mass spectrum tends to be stable when
$N^3$ is larger than $2^3$. Therefore, we expand the wave functions
of the tetraquark states with $2^3$ Gaussian basis in the following
calculation.

\subsection{A tetraquark state $QQ\bar Q'\bar Q'$ with $J^{PC}=0^{++}$}
A tetraquark state $QQ\bar Q'\bar Q'$ with $J^{PC}=0^{++}$ contains
two color-flavor-spin configurations $\chi_1$ and $\chi_2$ as listed
in Eq. (\ref{0p}). Its wave function reads
\begin{eqnarray}
&&\psi_{JJ_{z}}^{II_{z}}=\sum_{\alpha_{1}}A_{\alpha_{1}}\phi_{\alpha_{1}}\chi_{1}+\sum_{\alpha_{2}}B_{\alpha_{2}}\phi_{\alpha_{2}}\chi_{2}\nonumber\\
&&=\sum_{\alpha_{1}}A_{\alpha_{1}}\phi_{\alpha_{1}}(\beta_{a},\beta_{b},\beta)|(QQ)_{\bar 3_c}(\bar Q\bar Q)_{3_c}\rangle\nonumber\\
&&+\sum_{\alpha_{2}}B_{\alpha_{2}}\phi_{\alpha_{2}}(\gamma_a,\gamma_b,\gamma)|(QQ)_{6_c}(\bar
Q\bar Q)_{\bar 6_c}\rangle, \label{0++state}
\end{eqnarray}
where $\alpha_{1,2}=\{n_a,l_a,n_b,l_b,N,L\}$, $\beta_{(a,b)}$ and
$\gamma_{(a,b)}$ are the oscillating parameters for the $\bar
3_c-3_c$ and $6_c-\bar 6_c$ tetraquark states. $A_{\alpha_1}$ and
$B_{\alpha_2}$ are the expanding coefficients.

At first, we do not consider the mixture between the $\bar 3_c-3_c$
and $6_c-\bar 6_c$ tetraquark states and solve the Sch\"odinger
equation with the variational method. We obtain their mass spectra
and display them in the left panel of Fig.~\ref{couple effects}.

For the $cc\bar{c}\bar{c}$ and $bb\bar{b}\bar{b}$ systems, the
$6_c-\bar 6_c$ states are located lower than the $\bar 3_c-3_c$ ones
as illustrated in Fig.~\ref{couple effects}. In the OGE model, the
interactions between the two quarks within a color-sextet diquark
are repulsive due to the color factor in Table~\ref{color}, while
those in the $\bar 3_c$ one is attractive. However, the interactions
between the $6_c$ diquark and $\bar 6_c$ antidiquark are attractive
and much stronger than that between the $\bar 3_c$ diquark and $3_c$
antidiquark. There exists a $6_c-\bar 6_c$ tetraquark state, if the
attraction between diquark and antidiquark wins against the
repulsion within the diquak (antidiquark). If the attractive
potentials are strong enough, the $6_c-\bar 6_c$ state stays even
lower than the ${\bar 3_c}-{3_c}$ one. That is what happens to the
$cc\bar{c}\bar{c}$, $bb\bar{b}\bar{b}$ tetraquark states with
$J^{PC}=0^{++}$ in the two quark models. For the $bb\bar{c}\bar{c}$
($cc\bar b\bar b$) state, the $6_c-\bar 6_c$ state is lower in model
I, while the $\bar {3}_c-3_c$ state is lower in model II.

\begin{figure*}[htbp]
\centering \subfigure[~$cc\bar c\bar c$]{
\begin{minipage}[t]{0.323\linewidth}
\centering
\includegraphics[width=1\textwidth]{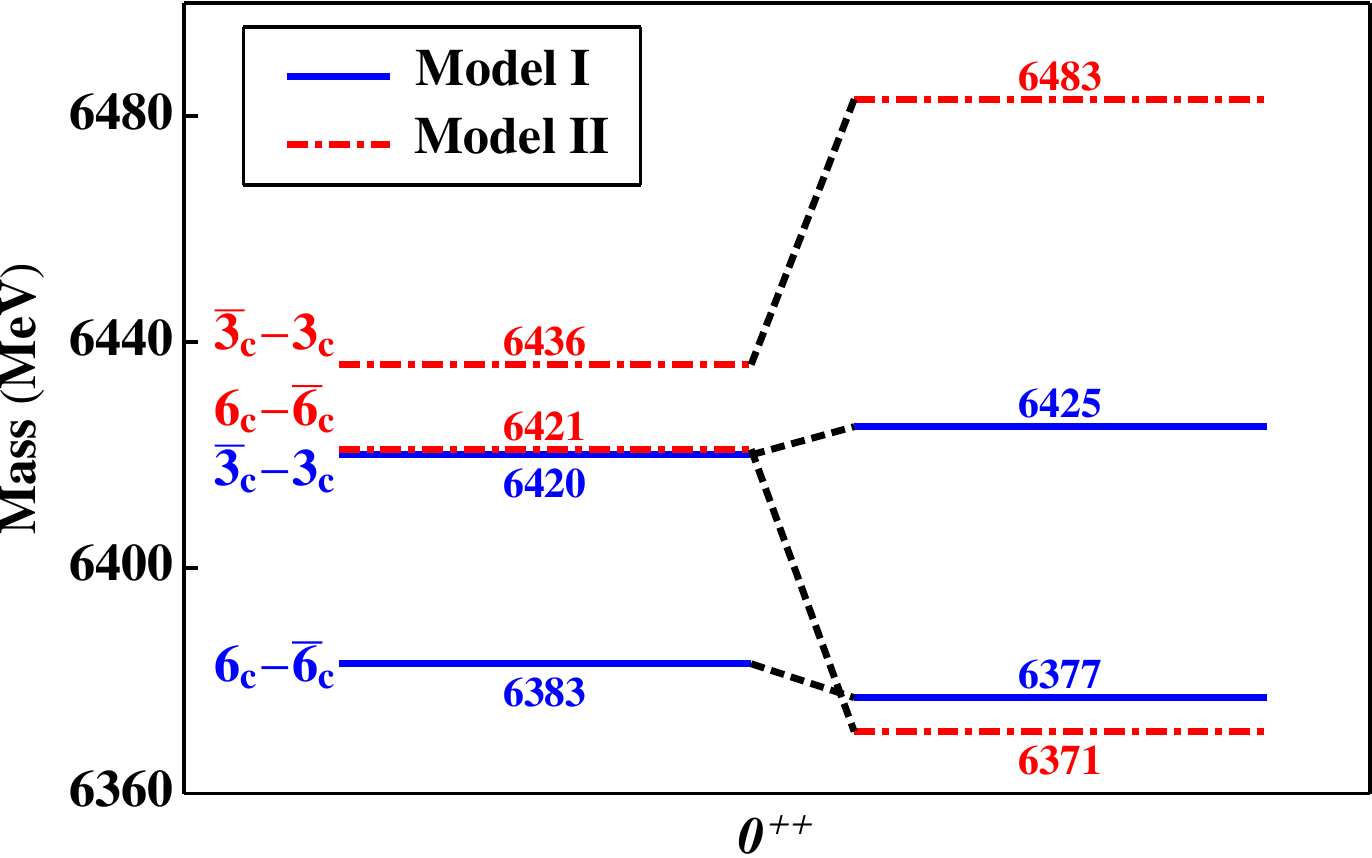}

\end{minipage}%
}%
\subfigure[~$bb\bar b\bar b$]{
\begin{minipage}[t]{0.33\linewidth}
\centering
\includegraphics[width=1\textwidth]{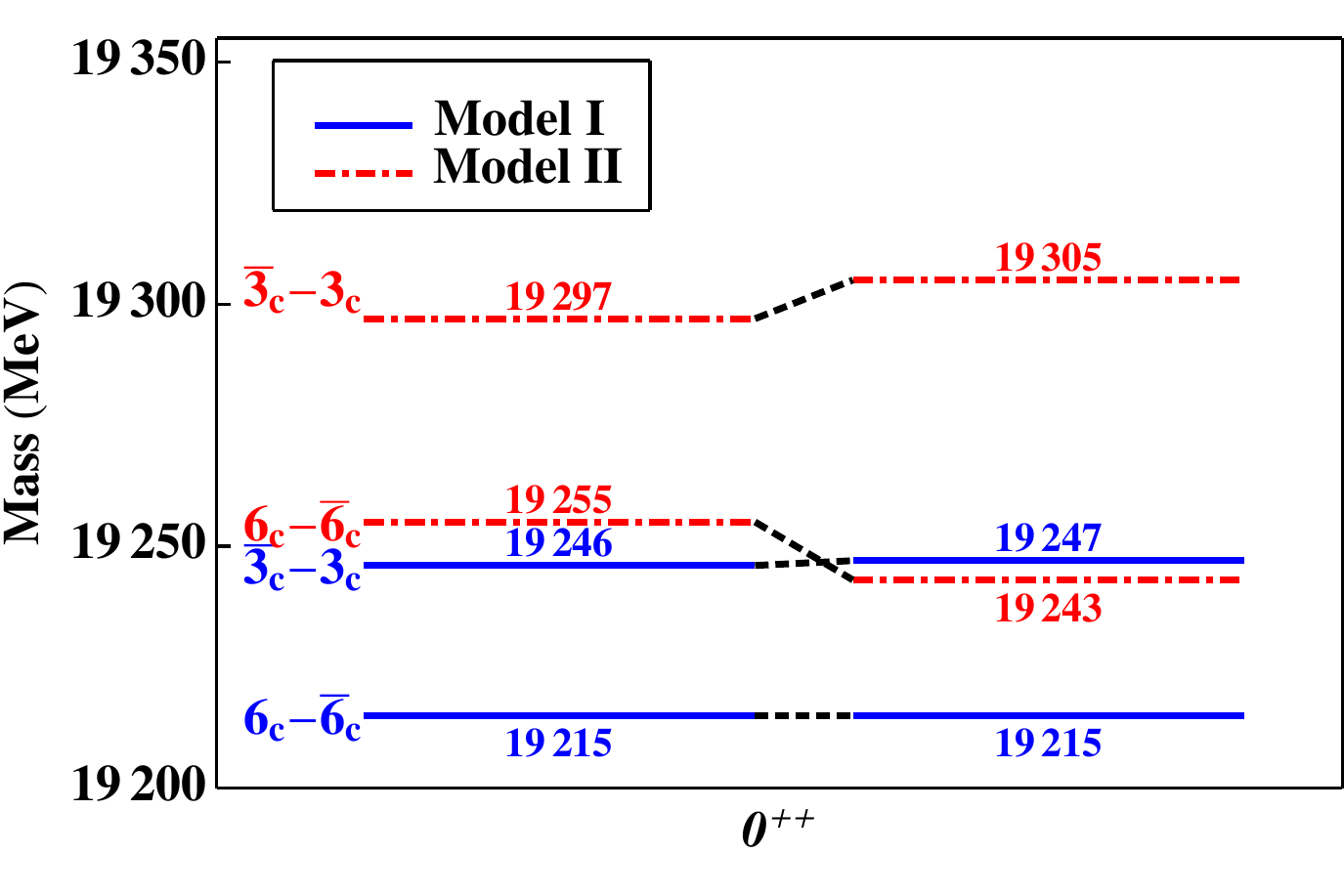}

\end{minipage}%
}%
\subfigure[~$bb\bar c\bar c$ ($cc\bar b\bar b$)]{
\begin{minipage}[t]{0.33\linewidth}
\centering
\includegraphics[width=1\textwidth]{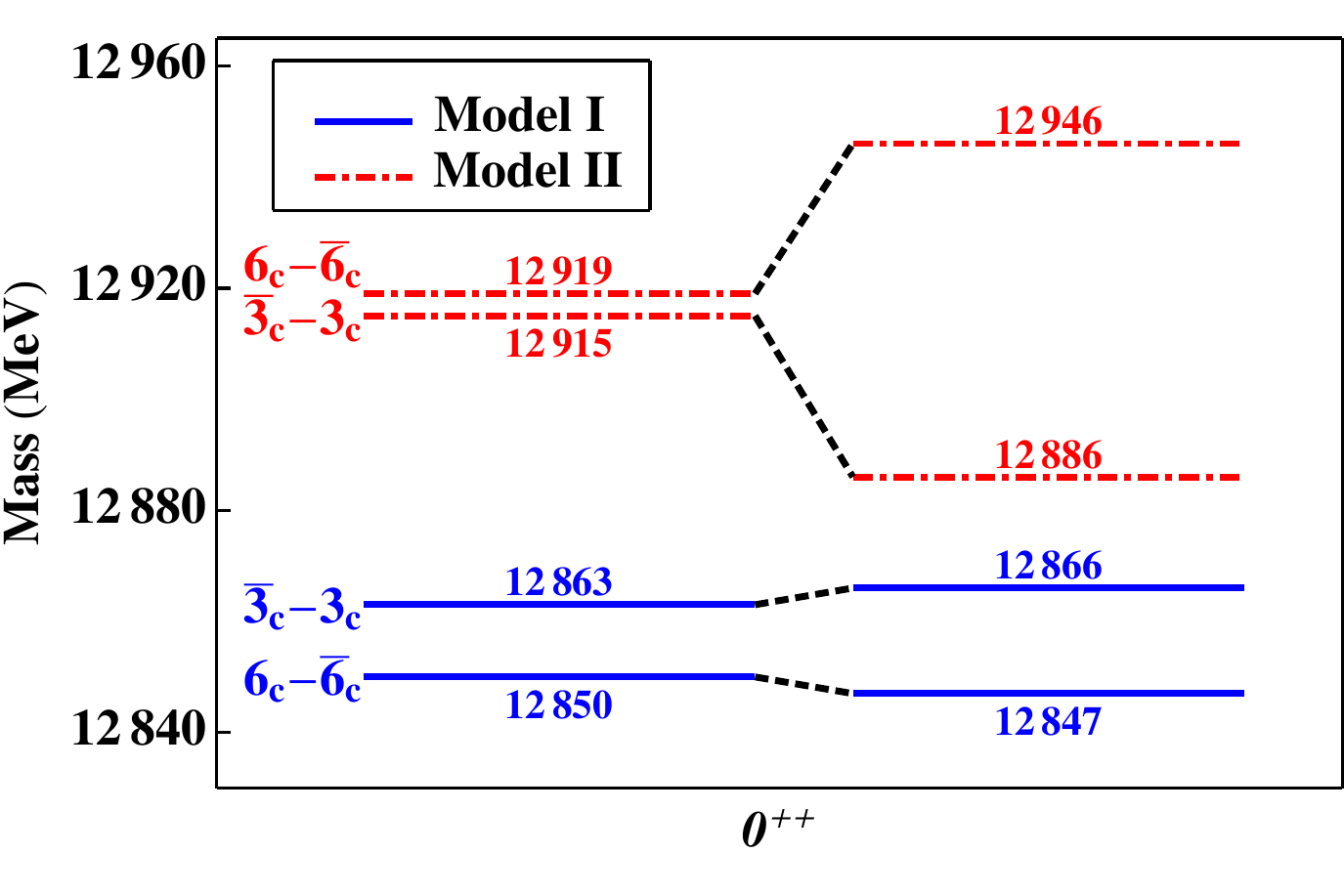}
\end{minipage}
}%
\centering \caption{ The mass spectrum of the $0^{++}$ tetraquark
states $QQ\bar Q'\bar Q'$ without and with the coupling between the
$\bar 3_c-3_c$ and $6_c-\bar 6_c$ configurations. The blue lines and
red dotted dashed lines represent the results in model I and II,
respectively. In every diagrams, the left half and the right half
are the mass spectrum without and with mixing between $\bar 3_c-3_c$
and $6_c-\bar 6_c$ configurations, respectively. The corresponding
states are connected by the black dashed lines. } \label{couple
effects}
\end{figure*}

\begin{table*}
\renewcommand\arraystretch{1.5}
\caption{The mass spectra of $cc\bar c\bar c$, $bb\bar b\bar b$, and
$bb\bar c\bar c$ ($\bar b\bar b cc$) tetraquark states with
$J^{PC}=0^{++}$. $\beta_{(a,b)}$ and $\gamma_{(a,b)}$ represent the
oscillating parameters of the $\bar 3_c-3_c$ and $6_c-\bar 6_c$
tetraquark states, respectively. }\label{0++} \centering
 \setlength{\tabcolsep}{1mm}
\begin{tabular}{l|cccc|cccc}
\toprule[1pt]\toprule[1pt] $J^{PC}=0^{++}$ & Model I & M [GeV] &
$\bar{3}_{c}\otimes3_{c}$ & $6_{c}\otimes\bar{6}_{c}$ & Model II & M
[GeV] & $\bar{3}_{c}\otimes3_{c}$ & $6_{c}\otimes\bar{6}_{c}$
\tabularnewline \midrule[1pt] \multirow{2}{*}{$cc\bar{c}\bar{c}$} &
$\beta_{a}=\beta_{b}=0.4$, $\beta=0.6$ & $6.377$ & $11\%$ & $89\%$ &
$\beta_{a}=\beta_{b}=0.5$, $\beta=0.7$ & $6.371$ & $43\%$ &
$57\%$\tabularnewline
 & $\gamma_{a}=\gamma_{b}=0.4$, $\gamma=0.7$ & $6.425$ & $89\%$ & $11\%$ & $\gamma_{a}=\gamma_{b}=0.5$, $\gamma=0.8$ & $6.483$ & $57\%$ & $43\%$\tabularnewline
 \midrule[1pt]
\multirow{2}{*}{$bb\bar{b}\bar{b}$} & $\beta_{a}=\beta_{b}=0.7$,
$\beta=0.9$ & $19.215$ & $1\%$ & $99\%$ & $\beta_{a}=\beta_{b}=0.9$,
$\beta=1.1$ & $19.243$ & $17\%$ & $83\%$\tabularnewline
 & $\gamma_{a}=\gamma_{b}=0.7$, $\gamma=0.9$ & $19.247$ & $99\%$ & $1\%$ & $\gamma_{a}=\gamma_{b}=0.8$, $\gamma=1.2$ & $19.305$ & $83\%$ & $17\%$\tabularnewline
 \midrule[1pt]
\multirow{2}{*}{$bb\bar{c}\bar{c}$} &
$\beta_{a}=0.6,\beta_{b}=0.5,\beta=0.7$ & $12.847$ & $14\%$ & $86\%$
& $\beta_{a}=0.7,\beta_{b}=0.5,\beta=0.8$ &$12.886$   &$53\%$ &$
47\%$ \tabularnewline
 & $\gamma_{a}=0.6,\gamma_{b}=0.4,\gamma=0.9$ & $12.866$ & $86\%$ & $14\%$ & $\gamma_{a}=0.7,\gamma_{b}=0.5,\gamma=0.9$ &$12.946$ &$47\%$ &$53\%$
\tabularnewline
 \bottomrule[1pt]\bottomrule[1pt]
\end{tabular}
\end{table*}

In general, a tetraquark state is a mixture of the $\bar 3_c-{3_c}$
and ${6_c}-{\bar 6_c}$ states as illustrated in
Eq.~(\ref{0++state}). With the couple-channel effects of the $\bar
3_c-3_c$ and $6_c-\bar 6_c$ color configurations, we obtain the mass
spectrum of the $0^{++}$ states and list them in Table~\ref{0++}.
The spectra obtained with $\bar 3_c-{3_c}$ and ${6_c}-{\bar 6_c}$
mixing are given in Fig.~\ref{couple effects}. The mixing effect
will pull down the lower state and raise the higher state. The two
quark models lead to similar mass spectra for the $cc\bar c\bar c$,
$bb\bar b\bar b$, and $bb\bar c\bar c$ ($cc\bar b\bar b$) tetraquark
states with the differences up to tens of MeV. However, the
proportions of the components in the two quark models are quite
different. The mixing between the $\bar 3_c-3_c$ and $6_c-\bar 6_c$
states are more stronger in model II. The reasons are explained as
follows.

In model I and model II, we find that only the hyperfine
interactions contribute to the couple-channel effects of the ${\bar
3_c}-{3_c}$ configuration and the ${6_c}-{\bar 6_c}$ one, while the
contributions from the confinement and Coulomb potentials vanish. We
illustrate the underlying dynamics as follows. The matrices of
$h_{12}$ and $h_{34}$ are diagonal due to the orthogonality of the
wave functions of different configurations. However, the
$V_{\text{coul}}+V_{\text{linear}}+V_{\text{hyp}}$ in $V_{I}$, which
describes the interactions between the diquark and antidiquark, may
result in the couple-channel effects of different configurations.
For an S-wave tetraquark state with two identical quarks
(antiquarks), such as $QQ\bar Q_1\bar Q_2$ ($Q_1Q_2\bar Q\bar Q$),
the spin wave functions of different possible configurations are
orthogonal, which is constrained by the Fermi statistic. Since the
OGE Coulomb and linear confinement potentials do not contain spin
operators, they do not contribute to the couple-channel effects due
to the orthogonality of the spin wave functions. And only the
hyperfine potential contributes. That is what happens to the $QQ\bar
Q'\bar Q'$ state in this work.

For a tetraquark state without identical quarks and antiquarks,
i.e., $Q_{1}Q_{2}\bar{Q}_{3}\bar{Q}_{4}$ ($Q_1\neq Q_2$ and $Q_3\neq
Q_4$), the spin wave functions of different configurations may be
the same. The four quarks form a color singlet state and the color
matrix element is
 \begin{eqnarray}
&& (\sum^4_n\mathbf \lambda_n)^2|\chi_{i,j} \rangle =0.%\nonumber \\
\end{eqnarray}
where $\chi_{i}$ and $\chi_j$ represent two different color
configurations and they are the eigenvectors of $\mathbf
\lambda_1+\mathbf \lambda_2$ and $\mathbf \lambda_3+\mathbf
\lambda_4$. Considering their orthogonality, one obtains
 \begin{eqnarray}
&& \langle \chi_i|(\mathbf \lambda_1+\mathbf \lambda_2)^2|\chi_j \rangle =0,\nonumber \\
&& \langle \chi_i|(\mathbf \lambda_3+\mathbf \lambda_4)^2|\chi_j \rangle =0. %\nonumber \\
\end{eqnarray}
Then the color factors of the (13), (14), (23), and (24) pairs of
quarks cancel out,
 \begin{eqnarray}
 && \langle \chi_i|(\lambda_1+\lambda_2)(\lambda_3+\lambda_4)|\chi_j\rangle=0.
\end{eqnarray}
Moreover, if the coupling constants are the same for the four quark
pairs, the contributions from the OGE Coulomb and the linear
confinement potentials will cancel out completely. In model I, the
contributions from the color interactions do not cancel out exactly
due to different $\alpha_s$. However, partial cancellations are
still expected. In model II, the OGE Coulomb and linear confinement
potentials do not depend on the mass of the interacting quarks.
Thus, the couple-channel effects arising from the OGE Coulomb and
linear confinement potentials cancel out. The mixing between
different color-flavor-spin configurations only comes from the
hyperfine potential, which is inversely proportional to the
interacting quark mass. Thus, the mixing in the $cc\bar{c}\bar{c}$
state is generally larger than that in the $bb\bar b\bar b$ state.

In model II, all the flavor dependence is packaged into the
hyperfine interaction, which is different from model I. The
hyperfine interaction in model II should play a more important role
than that in model I. Therefore, the couple-channel effect in model
II is stronger as illustrated in Fig.~\ref{couple effects}.

In model II, since the $r_0$ in the hyperfine interaction is the
function of the reduced mass between the two quarks, its value for
$b\bar c$ is in proximity to that of $c \bar c$. Then, the mixing in
$cc\bar c\bar c$ and $bb\bar c\bar c$ are similar as illustrated in
Table~\ref{0++}. One may wonder the additional dependence of the
mixing on the number of the expanding basis. For instance, when we
use $2\times3^{3}$ bases to expand the wave function of the $cc\bar
c\bar c$ state in model I, we find there are $11.4\%$ $\bar 3_c-3_c$
and $88.6\%$ $6_c-\bar 6_c$ components in the tetraquark state. The
percents change slightly with the number of the basis.

In Ref.~\cite{Liu:2019zuc}, the authors pointed out that the state
$|(QQ)_{\bar 3_c}(\bar Q \bar Q)_{3_c}\rangle$ $(Q=c,b)$ is located
lower than the $|(QQ)_{6_c}(\bar Q \bar Q)_{\bar 6_c}\rangle$ state,
which contradicts with our results. The inconsistency was due to
their use of particular wave functions. The authors used the same
oscillating parameters for the $\bar 3_c-3_c$ and $6_c-\bar 6_c$
states. Moreover, the oscillating parameters are proportional to the
reduced masses of the interacting quarks. With their wave function,
we reproduced their results. However, if we remove the two
constrains on the wave functions, we find the lowest state with a
dominant ${6_c}-{\bar 6_c}$ component as listed in
Table~\ref{comparison}, which is lower than that in
Ref.~\cite{Liu:2019zuc}.

\begin{table*}
\renewcommand\arraystretch{1.5}
\caption{The comparison of the mass spectra of $0^{++}$ $cc\bar
c\bar c$ and $bb\bar b\bar b$ from Ref.~\cite{Liu:2019zuc} and our
results using the same quark model. In the right table, we remove
the constrains on the wave functions used in
Ref.~\cite{Liu:2019zuc}. }\label{comparison} \centering
\begin{tabular}{l|cccc|cccc}
\toprule[1pt]\toprule[1pt] &
\multicolumn{4}{c|}{Ref.~\cite{Liu:2019zuc}} &
\multicolumn{4}{c}{without constrains}\tabularnewline \toprule[1pt]
$J^{PC}=0^{++}$ & $w=0.325$ & M [GeV] & $\bar{3}_{c}\otimes3_{c}$ &
$6_{c}\otimes\bar{6}_{c}$ & & M [GeV] & $\bar{3}_{c}\otimes3_{c}$ &
$6_{c}\otimes\bar{6}_{c}$\tabularnewline
 \midrule[1pt]
\multirow{2}{*}{$cc\bar{c}\bar{c}$}& $\beta_{a}=\beta_{b}=0.49$,
$\beta=0.69$ & $6470$ & $66\%$ & $34\%$ & $\beta_{a}=\beta_{b}=0.4$,
$\beta=0.6$ & $6417$ & $33\%$ & $67\%$\tabularnewline

 & $\gamma_{a}=\gamma_{b}=0.49$, $\gamma=0.69$ & $6559$ & $34\%$ & $66\%$ & $\gamma_{a}=\gamma_{b}=0.4$, $\gamma=0.7$ & $6509$ & $67\%$ & $33\%$\tabularnewline
 \midrule[1pt]
\multirow{2}{*}{$bb\bar{b}\bar{b}$} & $\beta_{a}=\beta_{b}=0.88$,
$\beta=1.24$ & $19268$ & $66\%$ & $34\%$ &
$\beta_{a}=\beta_{b}=0.7$, $\beta=0.9$ & $19226$ & $18\%$ &
$82\%$\tabularnewline

 & $\gamma_{a}=\gamma_{b}=0.88$, $\gamma=1.24$ & $19306$ & $34\%$ & $66\%$ & $\gamma_{a}=\gamma_{b}=0.7$, $\gamma=0.9$ & $19268$ & $82\%$ & $18\%$\tabularnewline
\bottomrule[1pt]\bottomrule[1pt]
\end{tabular}
\end{table*}

\subsection{The tetraquark states with $J^P=1^{+-}$ and $2^{++}$}
Constrained by the Fermi statistics, the tetraquark states $QQ\bar
Q'\bar Q'$ ($Q$ and $Q'$ may be the same flavors) with $J^P=1^{+-}$
and $2^{++}$ only contain one color component, i.e. $\bar 3_c-3_c$.
We list the mass spectra of the $S$-wave states and their radial
excitations in Table~\ref{j1j2}. The mass spectra in the two models
are quite similar to each other. The results from Model II are
slightly higher than those in Model I.

The tetraquark states with $J^P=1^{+-}$ and $2^{++}$ have the same
configurations except the total spin. Therefore, the mass difference
arises from the hyperfine potential, which is quite small compared
with the OGE Coulomb and linear confinement potentials. Thus, the
mass spectra of these two kinds of states are almost the same.

\begin{table*}
\renewcommand\arraystretch{1.5}
%\linespread{1.5}
\caption{The mass spectra of the $cc\bar c\bar c$, $bb\bar b\bar b$
and $bb\bar c\bar c $ states with $J^{PC}=1^{+-}$ and $2^{++}$ in
units of GeV. } \label{j1j2}
 \setlength{\tabcolsep}{2.5mm}
\begin{tabular}{c|cccc|cccc}
\toprule[1pt]\toprule[1pt]
 & Model I & $nS$ & $J^{PC}=1^{+-}$ & $J^{PC}=2^{++}$ & Model II & $nS$ & $J^{PC}=1^{+-}$ & $J^{PC}=2^{++}$\tabularnewline
 \bottomrule[1pt]
$cc\bar{c}\bar{c}$ & $\beta_{a}=0.4$ &$1S$ & $6.425$ & $6.432$ &
$\beta_{a}=0.5$ &$1S$& $6.450$ & $6.479$\tabularnewline

& $\beta_{b}=0.4$ &$2S$ & $6.856$ & $6.864$ & $\beta_{b}=0.5$ &$2S$
& $6.894$ & $6.919$\tabularnewline

& $\beta=0.6$ & $3S$ & $6.915$ & $6.919$ & $\beta=0.6$& $3S$ &
$7.036$ & $7.058$\tabularnewline \midrule[1pt]
 $bb\bar{b}\bar{b}$ & $\beta_{a}=0.7$&$1S$ & $19.247$ & $19.249$ & $\beta_{a}=1.0$ &$1S$ & $19.311$ & $19.325$\tabularnewline

& $\beta_{b}=0.7$&$2S$ & $19.594$ & $19.596$ & $\beta_{b}=1.0$ &$2S$
& $19.813$ & $19.823$\tabularnewline

& $\beta=0.9$ & $3S$& $19.681$ & $19.682$ & $\beta=1.1$ &$3S$ &
$20.065$ & $20.077$\tabularnewline

\midrule[1pt] $bb\bar{c}\bar{c}$ & $\beta_{a}=0.7$ & $1S$ & $12.864$
& $12.868$ & $\beta_{a}=0.7$ & $1S$ & $12.924$ &
$12.940$\tabularnewline

& $\beta_{b}=0.5$ & $2S$ & $13.259$ & $13.262$ & $\beta_{b}=0.5$ &
$2S$ & $13.321$ & $13.334$\tabularnewline & $\beta=0.7$ & $3S$ &
$13.297$ & $13.299$ & $\beta=0.7$ & $3S$ & $13.364$ &
$13.375$\tabularnewline \bottomrule[1pt] \bottomrule[1pt]
\end{tabular}
\end{table*}

\subsection{Disscussion}

A tetraquark state can be expressed in another set of color
representations as follows,
\begin{eqnarray}
    &&|(Q_{1}Q_{2})_{\bar{3}_{c}}(\bar{Q}_{3}\bar{Q}_{4})_{3_{c}}\rangle \nonumber \\
    &&=\sqrt{\frac{1}{3}}|(Q_{1}\bar{Q}_{3})_{1_{c}}(Q_{2}\bar{Q}_{4})_{1_{c}}\rangle-\sqrt{\frac{2}{3}}|(Q_{1}\bar{Q}_{3})_{8_{c}}(Q_{2}\bar{Q}_{4})_{8_{c}}\rangle \nonumber \\
    &&=-\sqrt{\frac{1}{3}}|(Q_{1}\bar{Q}_{4})_{1_{c}}(Q_{2}\bar{Q}_{3})_{1_{c}}\rangle+\sqrt{\frac{2}{3}}|(Q_{1}\bar{Q}_{4})_{8_{c}}(Q_{2}\bar{Q}_{3})_{8_{c}}\rangle,\nonumber \\
    &&|(Q_{1}Q_{2})_{6_{c}}(\bar{Q}_{3}\bar{Q}_{4})_{\bar{6}_{c}}\rangle \nonumber \\
    &&=\sqrt{\frac{2}{3}}|(Q_{1}\bar{Q}_{3})_{1_{c}}(Q_{2}\bar{Q}_{4})_{1_{c}}\rangle+\sqrt{\frac{1}{3}}|(Q_{1}\bar{Q}_{3})_{8_{c}}(Q_{2}\bar{Q}_{4})_{8_{c}}\rangle \nonumber \\
    &&=\sqrt{\frac{2}{3}}|(Q_{1}\bar{Q}_{4})_{1_{c}}(Q_{2}\bar{Q}_{3})_{1_{c}}\rangle+\sqrt{\frac{1}{3}}|(Q_{1}\bar{Q}_{4})_{8_{c}}(Q_{2}\bar{Q}_{3})_{8_{c}}\rangle. \nonumber \\
\end{eqnarray}
To investigate the inner structure of the tetraquark, we calculate
its proportions in the new set and the root mean square radii of the
state, which are listed in Table~\ref{proportion0++}. The ground
states contain the $8_c\otimes 8_c$ configuration. In model I, the
proportion of the $8_c\otimes 8_c$ configuration is considerable,
which supports that the solution is a confined state rather than a
scattering state of two mesons. In model II, though the $1_c\otimes
1_c$ configuration is dominant, the root mean square radii are of
the size of nucleons. Thus, they are also unlikely to be scattering
states.

\begin{figure*}[!tbp]
\centering {\subfigure[~Density distributions in the first Jacobi
coordinate.] {\includegraphics[width=0.43\textwidth]{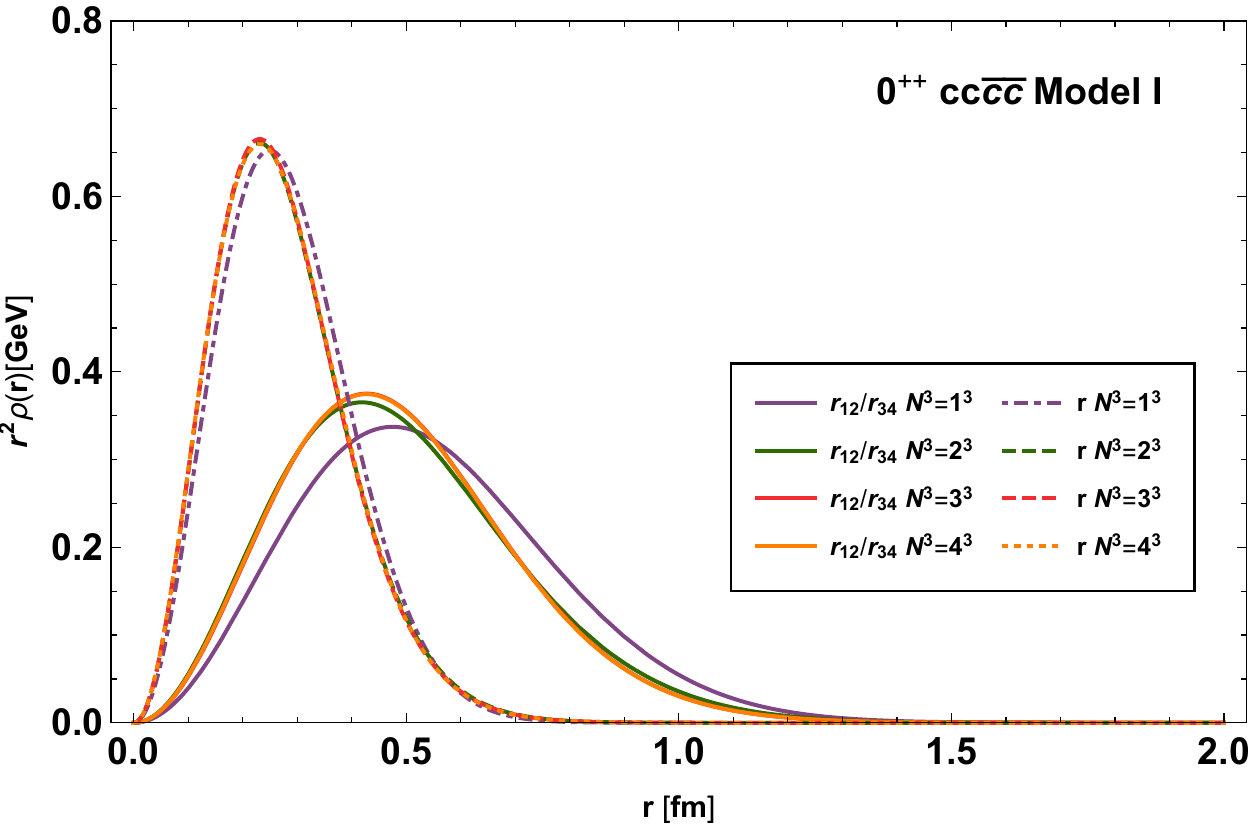}
\includegraphics[width=0.43\textwidth]{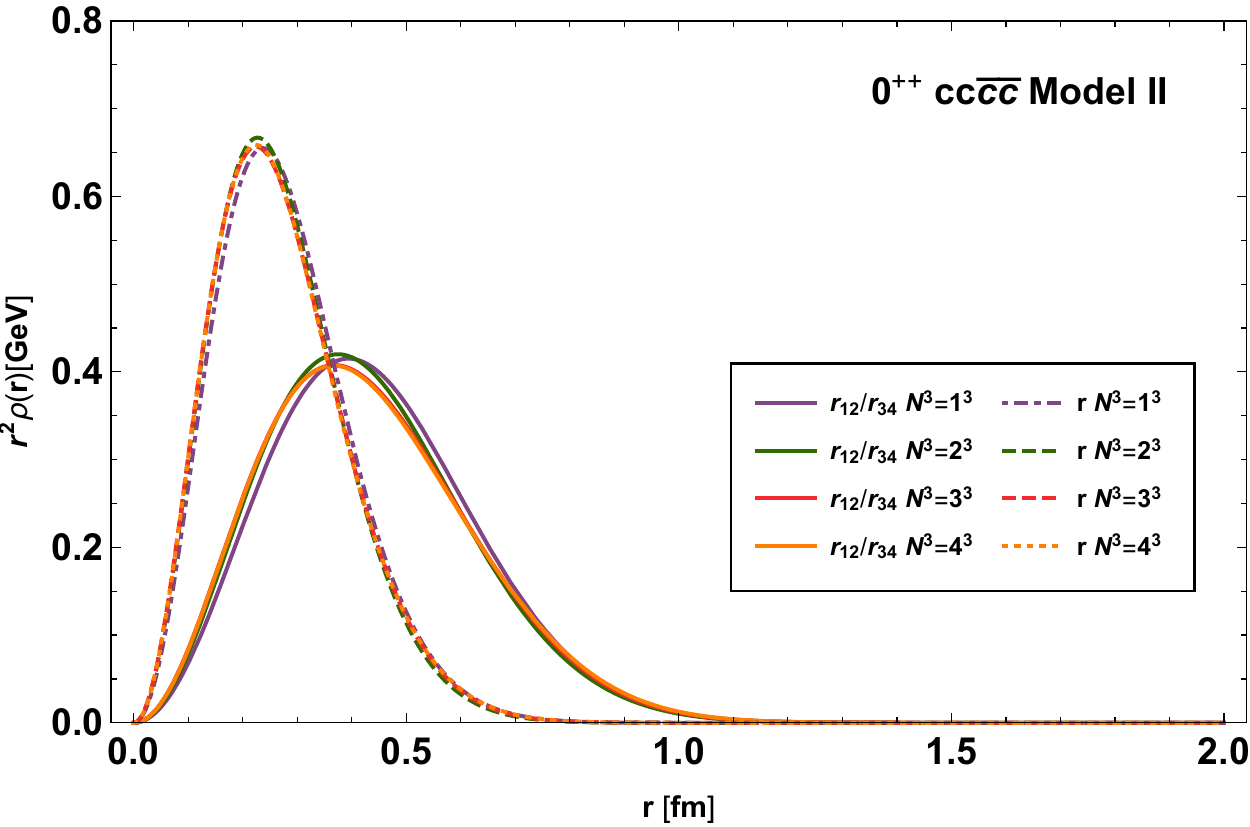}}}
{\subfigure[~Density distributions in the second Jacobi coordinate.]
{\includegraphics[width=0.43\textwidth]{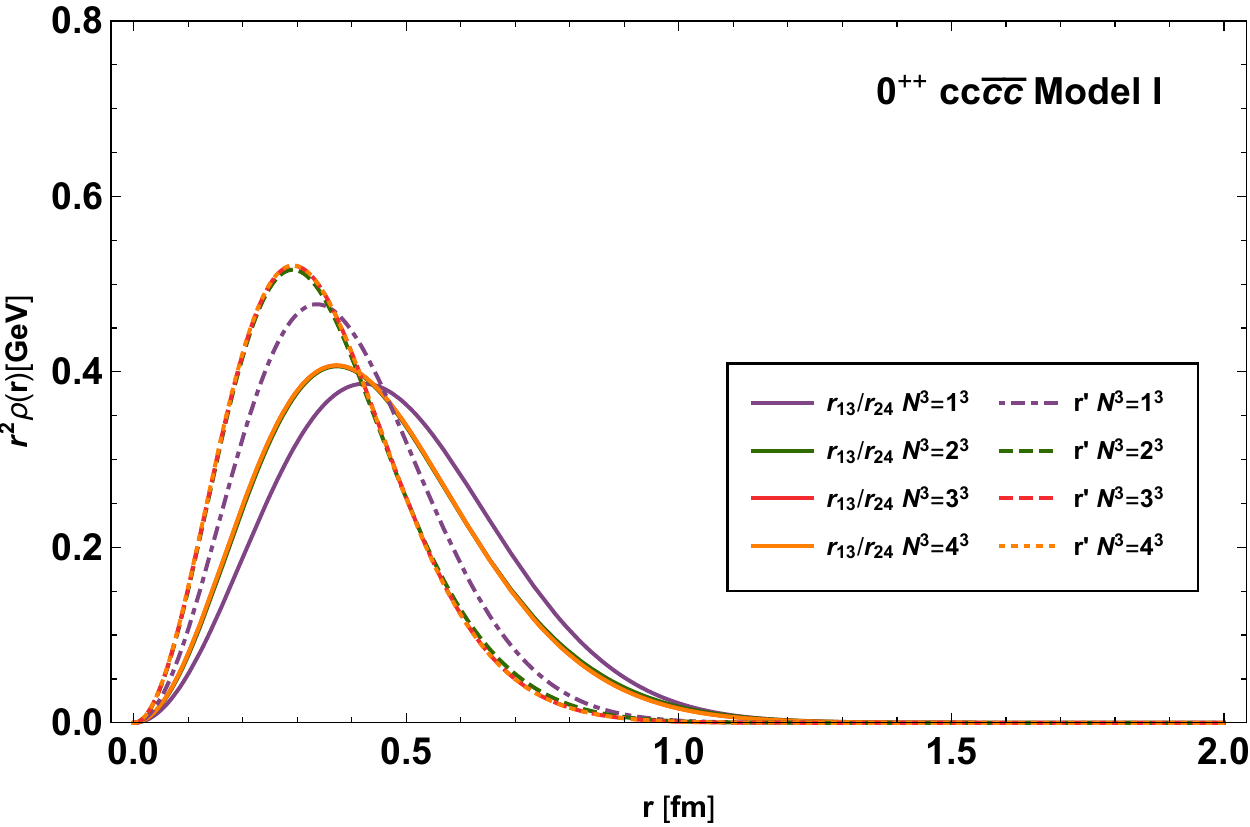}
\includegraphics[width=0.43\textwidth]{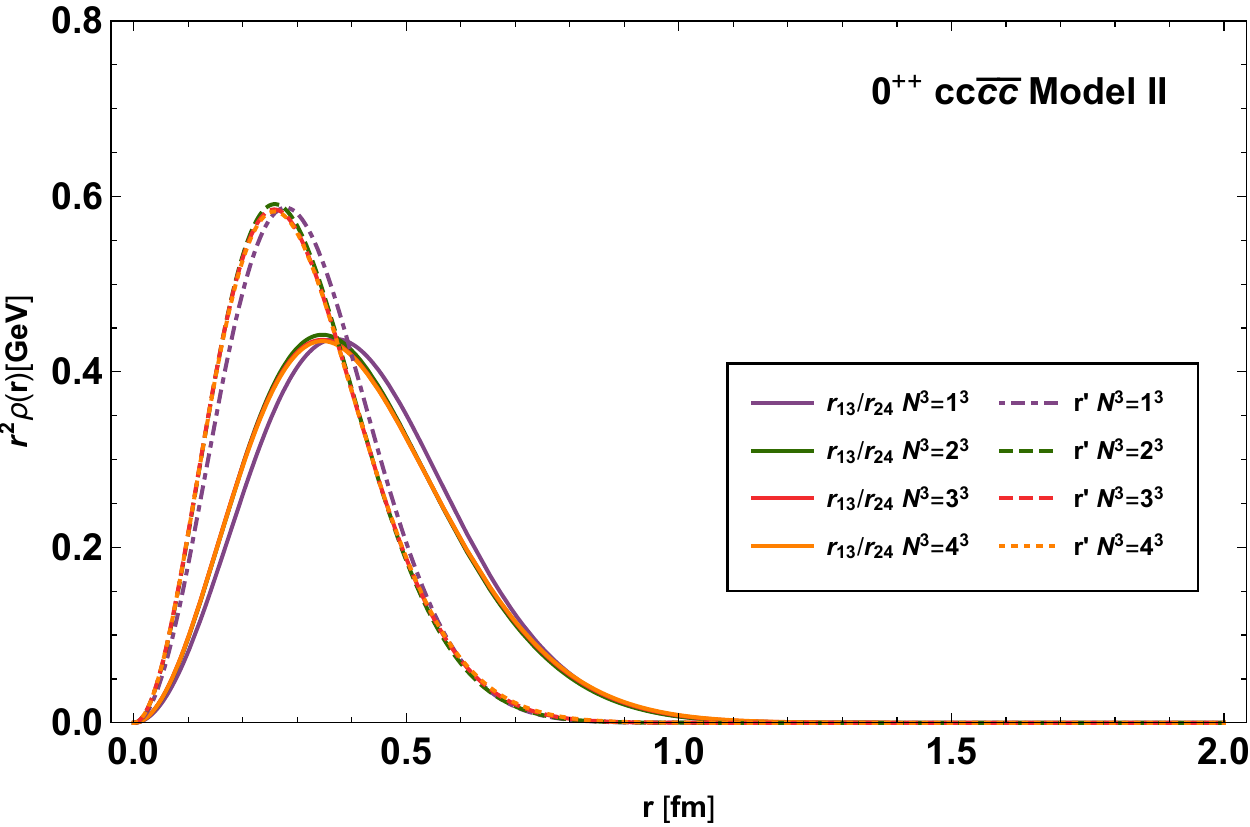}}}

\caption{The dependence of density distributions on the number of
the basis functions.} \label{dependence}
\end{figure*}

We also take the $cc\bar c\bar c$ as an example to study the density
distributions of $r^2\rho(r)$, $r^2\rho(r')$, $r^2_{12}
\rho(r_{12})$ and $r^2_{13} \rho(r_{13})$. The $\rho(r)$ and $
\rho(r_{12})$ are defined as follows,
\begin{eqnarray}
\ensuremath{\rho}(r)={\int}|\ensuremath{\psi}(r_{12},\,r_{34},\,r)|{}^{2}d\vec{r}_{12}d\vec{r}_{34}d\hat{\vec{r}},\nonumber \\
\rho(r_{12})=\ensuremath{\int}|\ensuremath{\psi}(r_{12},\,r_{34},\,r)|{}^{2}d\vec{r}d\vec{r}_{34}d\hat{\vec{r}}_{12}.
\end{eqnarray}
The definitions of the $\rho(r_{13})$ and $\rho(r')$ are similar.
The dependence of the density distributions on the extension of the
basis function is displayed in Fig.~\ref{dependence}. We find that
the distributions are confined in the spatial space and tend to be
stable with different number of the expanding basis, which indicates
the state may be a confined state instead of a scattering state.

\begin{table*}
\caption{The proportion of the color configurations and the root
mean square radii of the $cc\bar c\bar c$, $bb\bar b\bar b$, and
$bb\bar c\bar c$ ($\bar b\bar b cc$) tetraquark states with
$J^{PC}=0^{++}$. $\sqrt{\langle r_{ij}^{2}\rangle}$ and
$\sqrt{\langle r^{(')2}\rangle}$ are the root mean square radii
corresponding to the second Jacobi coordinate in Fig.~\ref{jac}.
}\label{proportion0++}
\begin{tabular}{c|c|c|c|c|c|c|c|c|c|c|c}
\toprule[1pt]\toprule[1pt] $J^{PC}=0^{++}$ &
\multicolumn{11}{c}{Model I}\tabularnewline \midrule[1pt]
$N^{3}=2^{3}$ & After mixing & $\bar{3}_{c}\otimes3_{c}$ &
$6_{c}\otimes\bar{6}_{c}$ & $1_{c}\otimes1_{c}$ &
$8_{c}\otimes8_{c}$ & $\sqrt{\langle r_{12}^{2}\rangle}$ fm &
$\sqrt{\langle r_{34}^{2}\rangle}$ fm & $\sqrt{\langle
r^{2}\rangle}$ fm & $\sqrt{\langle r_{13}^{2}\rangle}$ fm &
$\sqrt{\langle r_{24}^{2}\rangle}$ fm & $\sqrt{\langle
r'^{2}\rangle}$ fm\tabularnewline \midrule[1pt]
\multirow{1}{*}{$cc\bar{c}\bar{c}$} & $6.377$ & $11\%$ & $89\%$ &
$90\%$ & $10\%$ & \multicolumn{2}{c|}{$0.54$} & $0.30$ &
\multicolumn{2}{c|}{$0.49$} & $0.38$\tabularnewline \hline
\multirow{1}{*}{$bb\bar{b}\bar{b}$} & $19.215$ & $1\%$ & $99\%$ &
$75\%$ & $25\%$ & \multicolumn{2}{c|}{$0.35$} & $0.19$ &
\multicolumn{2}{c|}{$0.31$} & $0.25$\tabularnewline \hline
\multirow{1}{*}{$bb\bar{c}\bar{c}$} & $12.847$ & $14\%$ & $86\%$ &
$92\%$ & $8\%$ & $0.39$ & $0.50$ & $0.26$ &
\multicolumn{2}{c|}{$0.41$} & $0.32$\tabularnewline \midrule[1pt]
\multicolumn{12}{c}{Model II}\tabularnewline \midrule[1pt]
$N^{3}=2^{3}$ & After mixing & $\bar{3}_{c}\otimes3_{c}$ &
$6_{c}\otimes\bar{6}_{c}$ & $1_{c}\otimes1_{c}$ &
$8_{c}\otimes8_{c}$ & $\sqrt{\langle r_{12}^{2}\rangle}$ fm &
$\sqrt{\langle r_{34}^{2}\rangle}$ fm & $\sqrt{\langle
r^{2}\rangle}$ fm & $\sqrt{\langle r_{13}^{2}\rangle}$ fm &
$\sqrt{\langle r_{24}^{2}\rangle}$ fm & $\sqrt{\langle
r'^{2}\rangle}$ fm\tabularnewline \midrule[1pt]
\multirow{1}{*}{$cc\bar{c}\bar{c}$} & $6.371$ & $43\%$ & $57\%$ &
$97\%$ & $3\%$ & \multicolumn{2}{c|}{$0.47$} & $0.30$ &
\multicolumn{2}{c|}{$0.45$} & $0.33$\tabularnewline \hline
\multirow{1}{*}{$bb\bar{b}\bar{b}$} & $19.243$ & $17\%$ & $83\%$ &
$94\%$ & $6\%$ & \multicolumn{2}{c|}{$0.28$} & $0.17$ &
\multicolumn{2}{c|}{$0.26$} & $0.20$\tabularnewline \hline
\multirow{1}{*}{$bb\bar{c}\bar{c}$} & $12.886$ & $53\%$ & $47\%$ &
$93\%$ & $7\%$ & $0.32$ & \multicolumn{1}{c|}{$0.44$} & $0.26$ &
\multicolumn{2}{c|}{$0.37$} & $0.26$\tabularnewline
 \bottomrule[1pt]\bottomrule[1pt]
\end{tabular}
\end{table*}

\begin{figure*}[!tbp]
\centering {\subfigure[~$cc\bar c\bar c$]{
\begin{minipage}[t]{0.45\linewidth}
\centering
\includegraphics[width=1\textwidth]{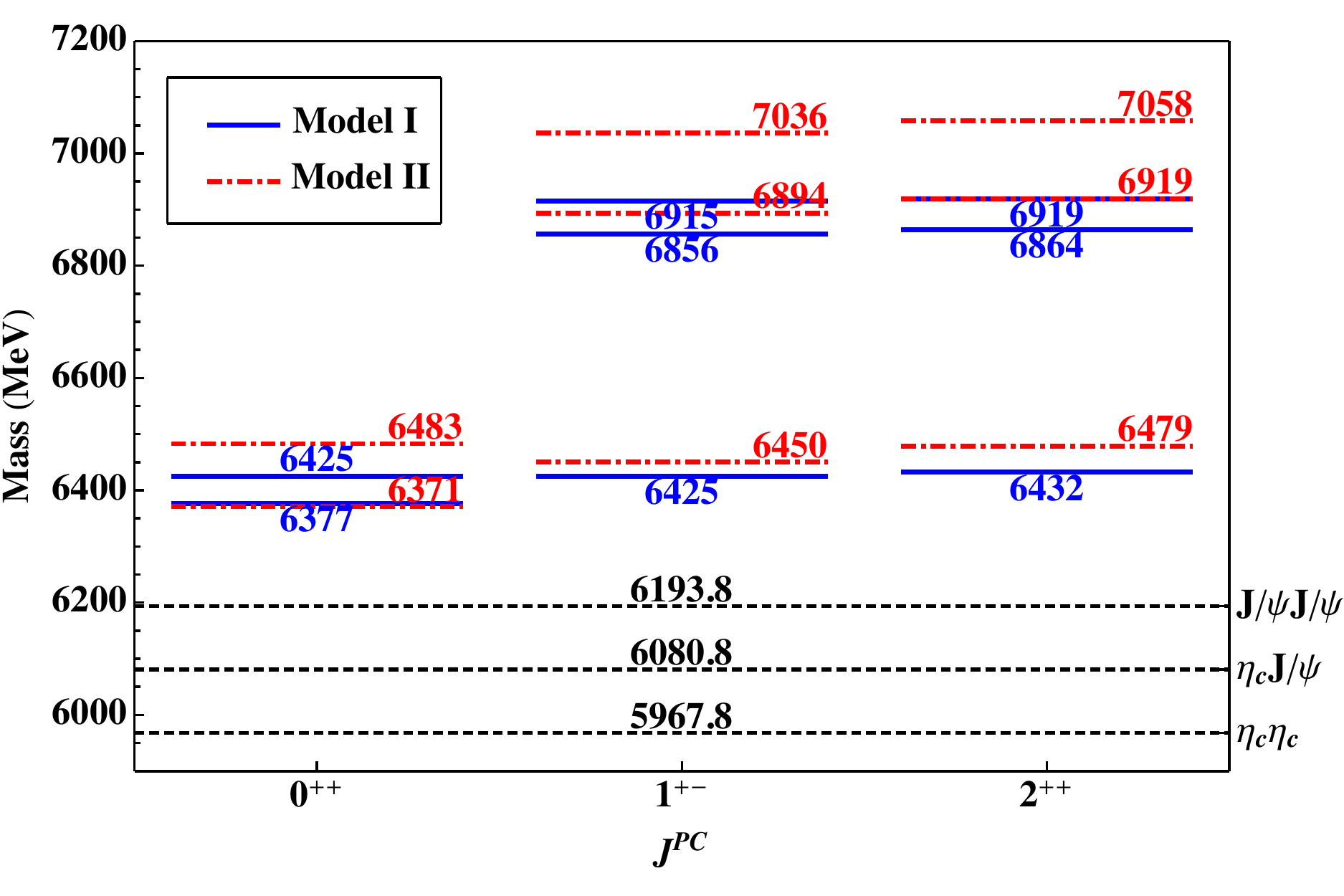}

\end{minipage}%
} \subfigure[~$bb\bar b\bar b$]
{\includegraphics[width=0.48\textwidth]{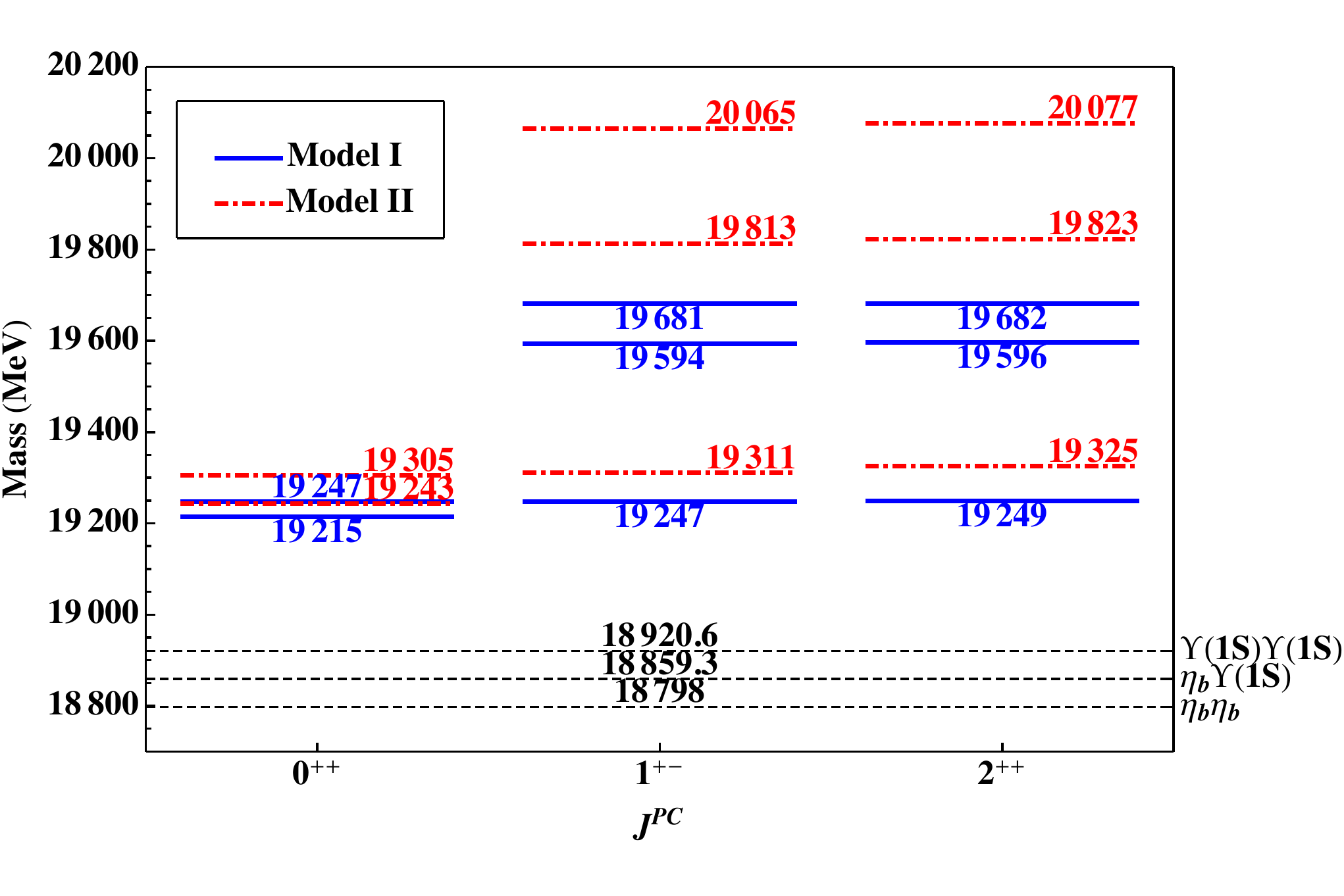}}
\subfigure[~$bb\bar c\bar c(cc\bar b\bar b)$.]
{\includegraphics[width=0.45\textwidth]{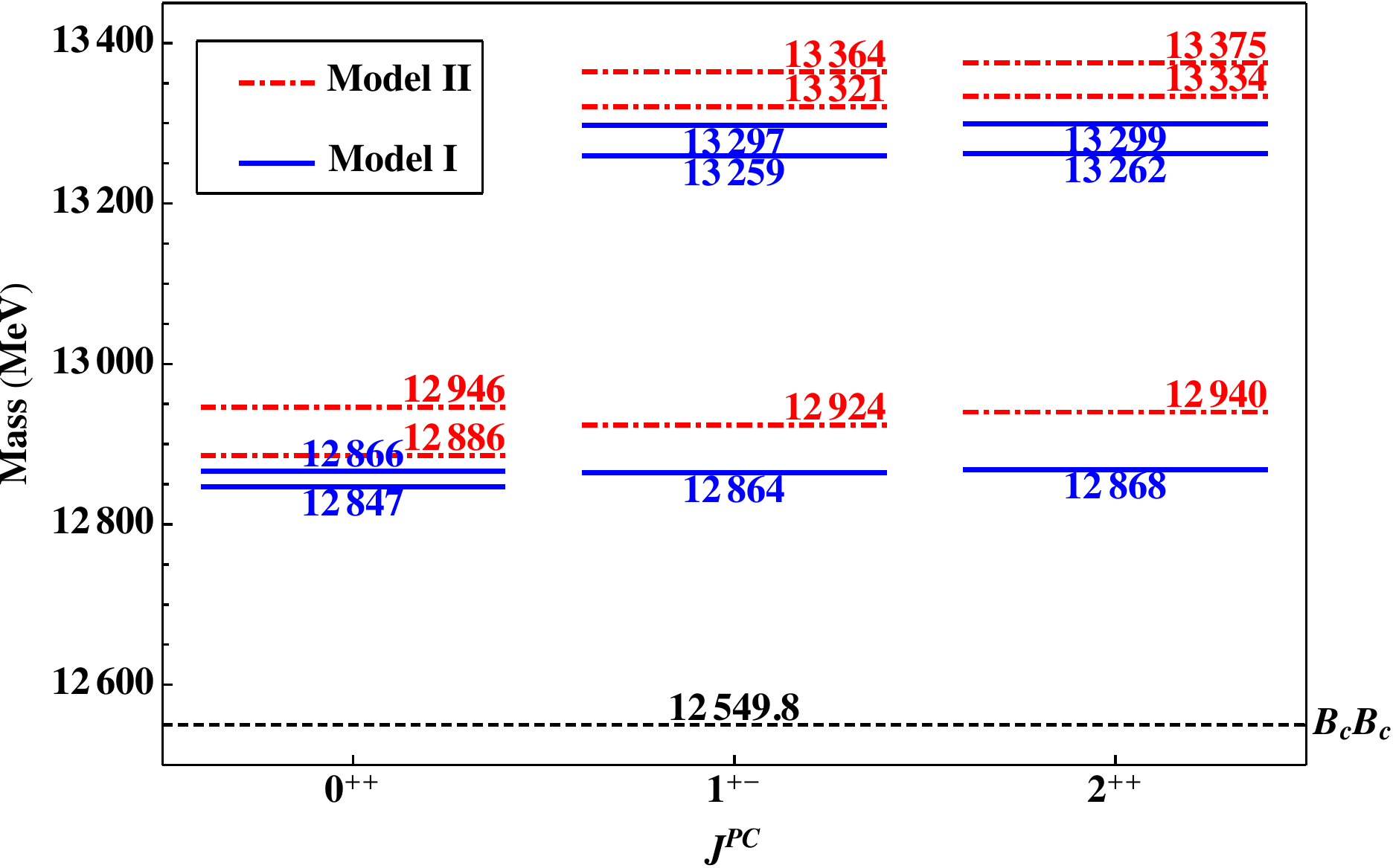}}} \caption{The
mass spectra of the $cc\bar c\bar c$, $bb\bar b\bar b$, and $bb\bar
c\bar c(cc\bar b\bar b)$ tetraquark states. The blue line and red
dotted dashed line represent the results in model I and II,
respectively. } \label{mass spectra}
\end{figure*}

We present the mass spectra of the tetraquark states and the mass
thresholds of possible scattering states in Fig.~\ref{mass spectra}.
As illustrated in this figure, the $bb\bar b \bar b$, $cc\bar c\bar
c$, and $bb\bar c\bar c$ states with $J^{PC}=0^{++}$ are the lowest
states. But they are still located above the corresponding
meson-meson mass thresholds, which indicates that there may not
exist bound states in the two quark models.

\begin{figure*}[!tbp]
\centering {\subfigure[The mass spectra of the tetraquark states
$QQ\bar Q\bar Q$ with $J^{PC}=0^{++}$.]
{\includegraphics[width=0.43\textwidth]{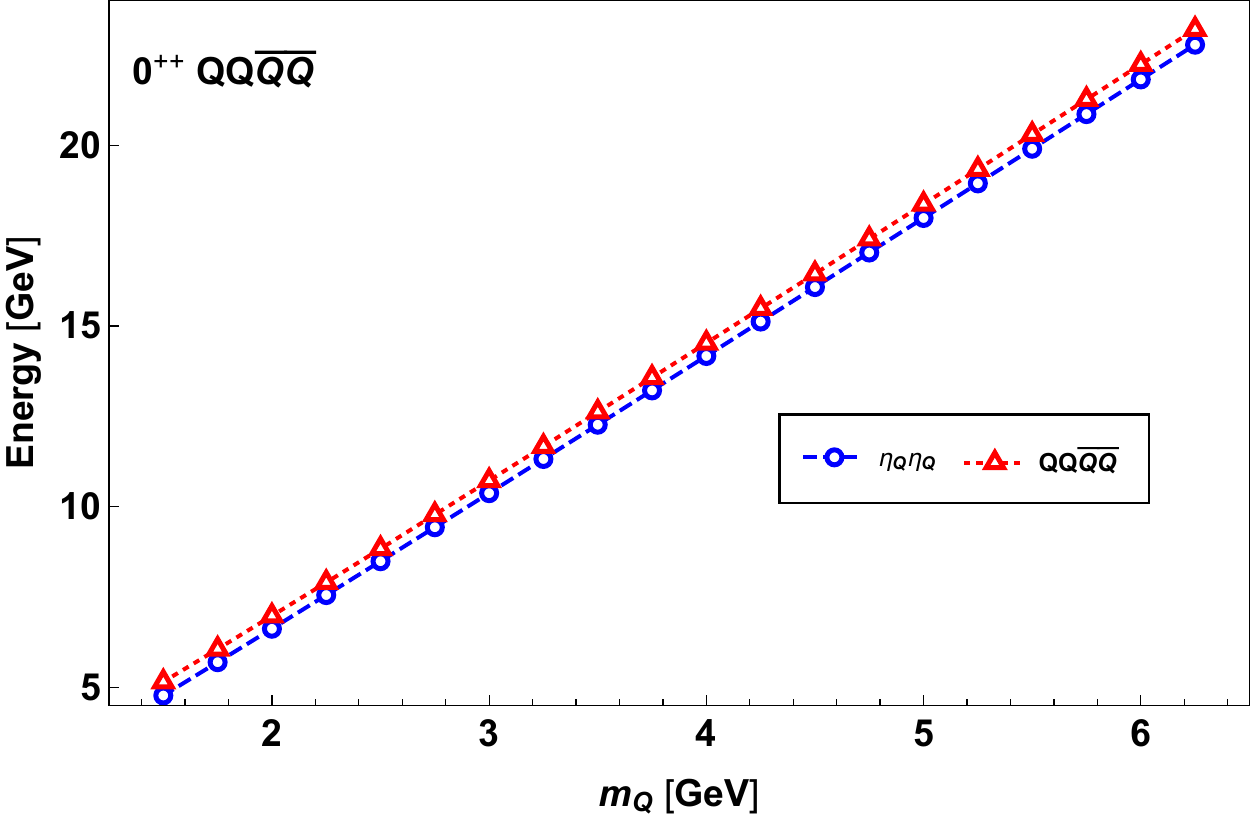}}
\subfigure[The mass difference between the tetraquark states and the
mass threshold of $\eta_Q\eta_Q$.]
{\includegraphics[width=0.44\textwidth]{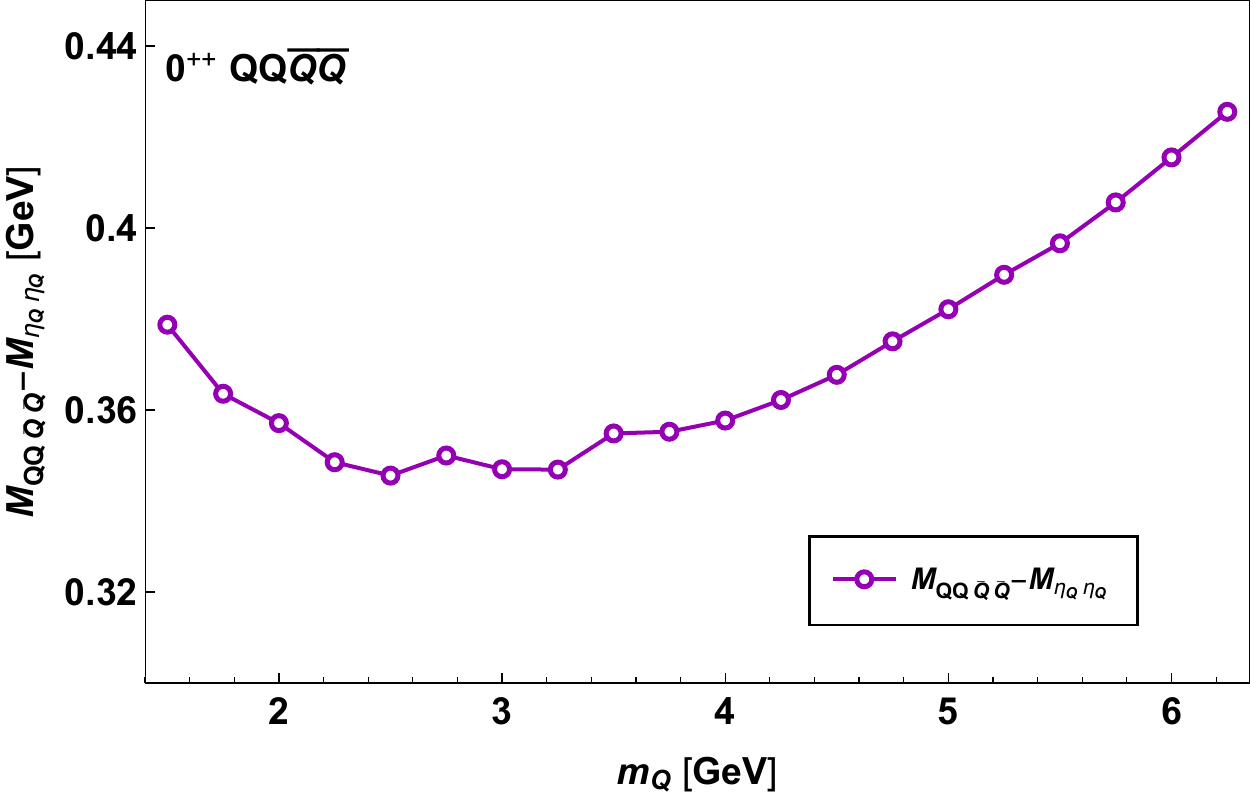}}} \caption{The
quark mass dependence of the $0^{++}$ tetraquark states $QQ\bar
Q\bar Q$ in model II. In this figure, we use the $\eta_Q$ to denote
the meson state $Q\bar Q$ with $J^{PC}=0^{-+}$.}
\label{massdependence}
\end{figure*}

We also investigate the constituent quark mass dependence of the
tetraquark spectra. We vary the quark mass and display the results
in Fig.~\ref{massdependence}. The figure shows that both the
tetraquark mass and the $\eta_Q\eta_Q$ threshold increase with the
quark mass. The $QQ\bar Q\bar Q$ is always located above the mass
thresholds of the $\eta_Q\eta_Q$ and no bound tetraquark states
exist.

\section{Summary}\label{sec3}

In this work, we have systematically calculated the mass spectra of
the tetraquark states $cc\bar c \bar c$, $bb\bar b\bar b$, and
$bb\bar c\bar c$ in two nonrelativistic quark models, which contain
the OGE Coulomb, linear confinement and hyperfine potentials.

For a $QQ\bar Q'\bar Q'$ ($Q$ and $Q'$ may be the same flavors)
state with $J^{PC}=0^{++}$, it can be formed by a $6_c$ diquark and
a $\bar 6_c$ antidiquark, or a $\bar 3_c$ diquark and a $3_c$
antidiquark. For the tetraquark states $cc\bar c \bar c$ and $bb\bar
b\bar b$, the $6_c-\bar 6_c$ states are located lower than the $\bar
3_c-3_c$ ones due to the strong attractions between the diquark and
the antidiquark. For the $bb\bar c\bar c$ ($cc\bar b\bar b$), the
mass of the $6_c-\bar 6_c$ state is lower than that of the $\bar
3_c-3_c$ one in the model I, while the $\bar 3_c-3_c$ one is lower
in the model II. Our calculation shows that the $6_c-\bar 6_c$ color
configuration is important and sometimes even dominant in the
formation of fully heavy tetraquark states. One should be cautious
about neglecting the $6_c-\bar 6_c$ color configurations in
calculating the tetraquark states.

The $6_c-\bar 6_c$ configuration couples with the $\bar 3_c-3_c$ one
through the interactions between the diquark and antidiquark. For a
$QQ\bar Q'\bar Q'$ state, we prove that only the hyperfine potential
contributes to the mixing between the two configurations, while the
contributions from the OGE Coulomb and the linear confinement
potentials cancel out exactly.
\begin{table*}
\centering \caption{The mass spectra (in units of GeV) of the
tetraquark states $cc\bar c\bar c$, $bb\bar b\bar b$, and $bb\bar
c\bar c$ in different frameworks. The $M^{1}_{th}$ and $M^{2}_{th}$
are the numerical results from the quark model I and II in this
work, respectively. } \label{summarize}
 \setlength{\tabcolsep}{1.7mm}
\begin{tabular}{l|cccccccccccccc}
\toprule[1pt]\toprule[1pt]
 & $J^{PC}$ & $M_{th}^{1}$ & $M_{th}^{2}$ &~\cite{Berezhnoy:2011xn}&~\cite{Karliner:2016zzc} &~\cite{Wu:2016vtq}&~\cite{Anwar:2017toa}&~\cite{Bai:2016int}&~\cite{Barnea:2006sd}&~\cite{Liu:2019zuc}&~\cite{Chen:2016jxd,Chen:2018cqz}\tabularnewline
\midrule[1pt]
 & \multirow{2}{*}{$0^{++}$} & $6.377$ & $6.371$ & \multirow{2}{*}{$5.966$ } & \multirow{2}{*}{$\ensuremath{6.192\pm0.025}$} & \multirow{2}{*}{$6.001$} & \multirow{2}{*}{$...$} & \multirow{2}{*}{$...$} & \multirow{2}{*}{$6.038$} & $6.470$ & \multirow{2}{*}{$6.44 \pm 0.15$} \tabularnewline
\multirow{3}{*}{$cc\bar{c}\bar{c}$} & & $6.425$ & $6.483$ & & & & &
& & $6.558$ & \tabularnewline \cline{2-12}
 & $1^{+-}$ & $6.425$ & $6.450$ & $6.051$ & $...$ & $6.109$ & $...$ & $...$ & $6.101$ & $6.512$ & $6.37\pm0.18$ \tabularnewline
 & $2^{++}$ & $6.432$ & $6.479$ & $6.223$ & $...$ & $6.166$ & $...$ & $...$ & $6.172$ & $6.534$ & $6.37 \pm 0.19$ \tabularnewline
\midrule[1pt] \multirow{4}{*}{$bb\bar{b}\bar{b}$} &
\multirow{2}{*}{$0^{++}$} & $19.215$ & $19.243$ &
\multirow{2}{*}{$18.754$} & \multirow{2}{*}{$18.826 \pm 0.025$} &
\multirow{2}{*}{$18.815$} & \multirow{2}{*}{$18.72 \pm 0.02$} &
\multirow{2}{*}{$18.69 \pm0.03$} & \multirow{2}{*}{$...$} & $19.268$
&\multirow{2}{*}{$18.45\pm0.15$} \tabularnewline

 & & $19.247$ & $19.305$ & & & & & & & $19.305$ & \tabularnewline
\cline{2-12}
 & $1^{+-}$ & $19.247$ & $19.311$ & $18.808$ & $...$ & $18.874$ & $...$ & $...$ & $...$ & $19.285$ &$ 18.32 \pm0.17$ \tabularnewline

 & $2^{++}$ & $19.249$ & $19.325$ & $18.916$ & $...$ & $18.905$ & $...$ & $...$ & $...$ & $19.295$ & $ 18.32 \pm 0.17$ \tabularnewline
\midrule[1pt] \multirow{4}{*}{$bb\bar{c}\bar{c}(cc\bar{b}\bar{b})$}
& \multirow{2}{*}{$0^{++}$} & $12.847$ & $12.886$ &
\multirow{2}{*}{$...$} & \multirow{2}{*}{$...$} &
\multirow{2}{*}{$12.571$} & \multirow{2}{*}{$...$} &
\multirow{2}{*}{$...$} & \multirow{2}{*}{$...$} & $12.935$
&\multirow{2}{*}{$...$}\tabularnewline

 & & $12.866$ & $12.946$ & & & & & & & $13.023$ & \tabularnewline
\cline{2-12}
 & $1^{+-}$ & $12.864$ & $12.924$ & $...$ & $...$ & $12.638$ & $...$ & $...$ & $...$ & $12.945$ & $...$\tabularnewline

 & $2^{++}$ & $12.868$ & $12.940$ & $...$ & $...$ & $12.673$ & $...$ & $...$ & $...$ & $12.956$ & $...$\tabularnewline
\bottomrule[1pt]\bottomrule[1pt]
\end{tabular}
\end{table*}

In Table~\ref{summarize}, we summarize our numerical results and
those from the CMI
model~\cite{Berezhnoy:2011xn,Karliner:2016zzc,Wu:2016vtq}, a
nonrelativistic effective field theory (NREFT) and a relativized
diquark and antidiquark model~\cite{Anwar:2017toa}, a diffusion
Monte-Carlo method~\cite{Bai:2016int}, a constituent quark model
with the hyperspherical formalism~\cite{Barnea:2006sd}, the
nonrelativistic potential model~\cite{Liu:2019zuc}, and the QCD sum
rule~\cite{Chen:2016jxd,Chen:2018cqz}. In this table, we notice that
the numerical results in the two nonrelativistic quark models are
similar to each other. The results show that the lowest states are
the ones with $J^{PC}=0^{++}$. These ground states are located about
$300\sim 450$ MeV above the lowest scattering states, which
indicates that there may not exist bound tetraquark states $cc\bar c
\bar c$, $bb\bar b\bar b$, and $bb\bar c\bar c$ $(cc\bar b\bar b)$
in the scheme of the two nonrelativistic quark models.

The parameters of the two quark models are determined by the meson
spectrum. The potentials in a four-body system may be slightly
different from those which are widely used in the conventional meson
and baryon systems. The different confinement mechanism may lead to
different spectra. For example, the three-body force arising from
the triple-gluon vertex may be non-negligible for the multi-quark
systems. In contrast, this force vanishes for the traditional $q\bar
q$ meson and $qqq$ baryons. The fully heavy tetraquark states can be
searched for at CMS, LHCb, and BelleII. More experimental data may
provide a deeper understanding of the interactions in the
multi-quark system.

\section*{ACKNOWLEDGMENTS}
G.J. Wang is very grateful to X. Z. Weng, X. L. Chen and W. Z. Deng
for very helpful discussions. We also thank Prof. Makoto Oka and
Prof. Emiko Hiyama for helpful suggestions. This project is
supported by the National Natural Science Foundation of China under
Grants 11575008, 11621131001 and 973 program.


\begin{thebibliography}{99}


%\cite{Choi:2007wga}
\bibitem{Choi:2007wga}
S.~K.~Choi {\it et al.} [Belle Collaboration],
%``Observation of a resonance-like structure in the $pi^\pm \psi^\prime$ mass distribution in exclusive $B \to K \pi^\pm \psi^\prime$ decays,''
Phys.\ Rev.\ Lett.\ {\bf 100}, 142001 (2008)
doi:10.1103/PhysRevLett.100.142001 [arXiv:0708.1790 [hep-ex]].
%%CITATION = doi:10.1103/PhysRevLett.100.142001;%%
%601 citations counted in INSPIRE as of 24 May 2019



%\cite{Aaij:2014jqa}
\bibitem{Aaij:2014jqa}
R.~Aaij {\it et al.} [LHCb Collaboration],
%``Observation of the resonant character of the $Z(4430)^-$ state,''
Phys.\ Rev.\ Lett.\ {\bf 112}, no. 22, 222002 (2014)
doi:10.1103/PhysRevLett.112.222002 [arXiv:1404.1903 [hep-ex]].
%%CITATION = doi:10.1103/PhysRevLett.112.222002;%%
%322 citations counted in INSPIRE as of 24 May 2019



%\cite{Chilikin:2013tch}
\bibitem{Chilikin:2013tch}
K.~Chilikin {\it et al.} [Belle Collaboration],
%``Experimental constraints on the spin and parity of the $Z$(4430)$^+$,''
Phys.\ Rev.\ D {\bf 88}, no. 7, 074026 (2013)
doi:10.1103/PhysRevD.88.074026 [arXiv:1306.4894 [hep-ex]].
%%CITATION = doi:10.1103/PhysRevD.88.074026;%%
%143 citations counted in INSPIRE as of 24 May 2019



%\cite{Chilikin:2014bkk}
\bibitem{Chilikin:2014bkk}
K.~Chilikin {\it et al.} [Belle Collaboration],
%``Observation of a new charged charmoniumlike state in $\bar{B}^0 â?J/ÏK^-Ï^+$ decays,''
Phys.\ Rev.\ D {\bf 90}, no. 11, 112009 (2014)
doi:10.1103/PhysRevD.90.112009 [arXiv:1408.6457 [hep-ex]].
%%CITATION = doi:10.1103/PhysRevD.90.112009;%%
%141 citations counted in INSPIRE as of 24 May 2019



%\cite{Ablikim:2013xfr}
\bibitem{Ablikim:2013xfr}
M.~Ablikim {\it et al.} [BESIII Collaboration],
%``Observation of a charged $(D\bar{D}^{*})^\pm$ mass peak in $e^{+}e^{-} \to \pi D\bar{D}^{*}$ at $\sqrt{s} =$ 4.26 GeV,''
Phys.\ Rev.\ Lett.\ {\bf 112}, no. 2, 022001 (2014)
doi:10.1103/PhysRevLett.112.022001 [arXiv:1310.1163 [hep-ex]].
%%CITATION = doi:10.1103/PhysRevLett.112.022001;%%
%251 citations counted in INSPIRE as of 24 May 2019



%\cite{Ablikim:2013wzq}
\bibitem{Ablikim:2013wzq}
M.~Ablikim {\it et al.} [BESIII Collaboration],
%``Observation of a Charged Charmoniumlike Structure $Z_c$(4020) and Search for the $Z_c$(3900) in $e^+e^- \to Ï^+Ï^-h_c$,''
Phys.\ Rev.\ Lett.\ {\bf 111}, no. 24, 242001 (2013)
doi:10.1103/PhysRevLett.111.242001 [arXiv:1309.1896 [hep-ex]].
%%CITATION = doi:10.1103/PhysRevLett.111.242001;%%
%327 citations counted in INSPIRE as of 24 May 2019



%\cite{Ablikim:2013mio}
\bibitem{Ablikim:2013mio}
M.~Ablikim {\it et al.} [BESIII Collaboration],
%``Observation of a Charged Charmoniumlike Structure in $e^+e^-$ â?$Ï^+Ï^-$ J/Ï at $\sqrt{s}$ =4.26 GeV,''
Phys.\ Rev.\ Lett.\ {\bf 110}, 252001 (2013)
doi:10.1103/PhysRevLett.110.252001 [arXiv:1303.5949 [hep-ex]].
%%CITATION = doi:10.1103/PhysRevLett.110.252001;%%
%656 citations counted in INSPIRE as of 24 May 2019



%\cite{Belle:2011aa}
\bibitem{Belle:2011aa}
A.~Bondar {\it et al.} [Belle Collaboration],
%``Observation of two charged bottomonium-like resonances in Y(5S) decays,''
Phys.\ Rev.\ Lett.\ {\bf 108}, 122001 (2012)
doi:10.1103/PhysRevLett.108.122001 [arXiv:1110.2251 [hep-ex]].
%%CITATION = doi:10.1103/PhysRevLett.108.122001;%%
%476 citations counted in INSPIRE as of 24 May 2019



%\cite{Adachi:2012cx}
\bibitem{Adachi:2012cx}
I.~Adachi {\it et al.} [Belle Collaboration],
%``Study of Three-Body Y(10860) Decays,''
arXiv:1209.6450 [hep-ex].
%%CITATION = ARXIV:1209.6450;%%
%92 citations counted in INSPIRE as of 24 May 2019



%\cite{Aaij:2015tga}
\bibitem{Aaij:2015tga}
R.~Aaij {\it et al.} [LHCb Collaboration],
%``Observation of $J/\psi p$ Resonances Consistent with Pentaquark States in $\Lambda_b^0 \to J/\psi K^- p$ Decays,''
Phys.\ Rev.\ Lett.\ {\bf 115}, 072001 (2015)
doi:10.1103/PhysRevLett.115.072001 [arXiv:1507.03414 [hep-ex]].
%%CITATION = doi:10.1103/PhysRevLett.115.072001;%%
%730 citations counted in INSPIRE as of 24 May 2019



%\cite{Aaij:2018bla}
\bibitem{Aaij:2018bla}
R.~Aaij {\it et al.} [LHCb Collaboration],
%``Evidence for an $\eta _c(1S) \pi ^-$ resonance in $B^0 \rightarrow \eta _c(1S) K^+\pi ^-$ decays,''
Eur.\ Phys.\ J.\ C {\bf 78}, no. 12, 1019 (2018)
doi:10.1140/epjc/s10052-018-6447-z [arXiv:1809.07416 [hep-ex]].
%%CITATION = doi:10.1140/epjc/s10052-018-6447-z;%%
%18 citations counted in INSPIRE as of 24 May 2019



%\cite{Chen:2016qju}
\bibitem{Chen:2016qju}
H.~X.~Chen, W.~Chen, X.~Liu and S.~L.~Zhu,
%``The hidden-charm pentaquark and tetraquark states,''
Phys.\ Rept.\ {\bf 639}, 1 (2016) doi:10.1016/j.physrep.2016.05.004
[arXiv:1601.02092 [hep-ph]].
%%CITATION = doi:10.1016/j.physrep.2016.05.004;%%
%407 citations counted in INSPIRE as of 24 May 2019



%\cite{Guo:2017jvc}
\bibitem{Guo:2017jvc}
F.~K.~Guo, C.~Hanhart, U.~G.~MeiÃner, Q.~Wang, Q.~Zhao and
B.~S.~Zou,
%``Hadronic molecules,''
Rev.\ Mod.\ Phys.\ {\bf 90}, no. 1, 015004 (2018)
doi:10.1103/RevModPhys.90.015004 [arXiv:1705.00141 [hep-ph]].
%%CITATION = doi:10.1103/RevModPhys.90.015004;%%
%235 citations counted in INSPIRE as of 24 May 2019



%\cite{Esposito:2016noz}
\bibitem{Esposito:2016noz}
A.~Esposito, A.~Pilloni and A.~D.~Polosa,
%``Multiquark Resonances,''
Phys.\ Rept.\ {\bf 668}, 1 (2017) doi:10.1016/j.physrep.2016.11.002
[arXiv:1611.07920 [hep-ph]].
%%CITATION = doi:10.1016/j.physrep.2016.11.002;%%
%191 citations counted in INSPIRE as of 24 May 2019



%\cite{Ali:2017jda}
\bibitem{Ali:2017jda}
A.~Ali, J.~S.~Lange and S.~Stone,
%``Exotics: Heavy Pentaquarks and Tetraquarks,''
Prog.\ Part.\ Nucl.\ Phys.\ {\bf 97}, 123 (2017)
doi:10.1016/j.ppnp.2017.08.003 [arXiv:1706.00610 [hep-ph]].
%%CITATION = doi:10.1016/j.ppnp.2017.08.003;%%
%133 citations counted in INSPIRE as of 24 May 2019



%\cite{Liu:2019zoy}
\bibitem{Liu:2019zoy}
Y.~R.~Liu, H.~X.~Chen, W.~Chen, X.~Liu and S.~L.~Zhu,
%``Pentaquark and Tetraquark states,''
doi:10.1016/j.ppnp.2019.04.003 arXiv:1903.11976 [hep-ph].
%%CITATION = doi:10.1016/j.ppnp.2019.04.003;%%
%17 citations counted in INSPIRE as of 24 May 2019



%\cite{Eichten:1978tg}
%\bibitem{Eichten:1978tg}
%E.~Eichten, K.~Gottfried, T.~Kinoshita, K.~D.~Lane and T.~M.~Yan,
%``Charmonium: The Model,''
%Phys.\ Rev.\ D {\bf 17}, 3090 (1978)
%Erratum: [Phys.\ Rev.\ D {\bf 21}, 313 (1980)].
%doi:10.1103/PhysRevD.17.3090, 10.1103/physrevd.21.313.2
%%CITATION = doi:10.1103/PhysRevD.17.3090, 10.1103/physrevd.21.313.2;%%
%1418 citations counted in INSPIRE as of 24 May 2019



%\cite{Eichten:1979ms}
%\bibitem{Eichten:1979ms}
%E.~Eichten, K.~Gottfried, T.~Kinoshita, K.~D.~Lane and T.~M.~Yan,
%``Charmonium: Comparison with Experiment,''
%Phys.\ Rev.\ D {\bf 21}, 203 (1980).
%doi:10.1103/PhysRevD.21.203
%%CITATION = doi:10.1103/PhysRevD.21.203;%%
%1691 citations counted in INSPIRE as of 24 May 2019



%\cite{Maiani:2004vq}
\bibitem{Maiani:2004vq}
L.~Maiani, F.~Piccinini, A.~D.~Polosa and V.~Riquer,
%``Diquark-antidiquarks with hidden or open charm and the nature of X(3872),''
Phys.\ Rev.\ D {\bf 71}, 014028 (2005)
doi:10.1103/PhysRevD.71.014028 [hep-ph/0412098].
%%CITATION = doi:10.1103/PhysRevD.71.014028;%%
%660 citations counted in INSPIRE as of 24 May 2019



%\cite{Ali:2011ug}
\bibitem{Ali:2011ug}
A.~Ali, C.~Hambrock and W.~Wang,
%``Tetraquark Interpretation of the Charged Bottomonium-like states $Z_b^{+-}(10610)$ and $Z_b^{+-}(10650)$ and Implications,''
Phys.\ Rev.\ D {\bf 85}, 054011 (2012)
doi:10.1103/PhysRevD.85.054011 [arXiv:1110.1333 [hep-ph]].
%%CITATION = doi:10.1103/PhysRevD.85.054011;%%
%93 citations counted in INSPIRE as of 24 May 2019



%\cite{Eichten:2017ffp}
\bibitem{Eichten:2017ffp}
E.~J.~Eichten and C.~Quigg,
%``Heavy-quark symmetry implies stable heavy tetraquark mesons $Q_iQ_j \bar q_k \bar q_l$,''
Phys.\ Rev.\ Lett.\ {\bf 119}, no. 20, 202002 (2017)
doi:10.1103/PhysRevLett.119.202002 [arXiv:1707.09575 [hep-ph]].
%%CITATION = doi:10.1103/PhysRevLett.119.202002;%%
%76 citations counted in INSPIRE as of 24 May 2019



%\cite{Chen:2016oma}
\bibitem{Chen:2016oma}
H.~X.~Chen, E.~L.~Cui, W.~Chen, X.~Liu and S.~L.~Zhu,
%``Understanding the internal structures of the $X(4140)$, $X(4274)$, $X(4500)$ and $X(4700)$,''
Eur.\ Phys.\ J.\ C {\bf 77}, no. 3, 160 (2017)
doi:10.1140/epjc/s10052-017-4737-5 [arXiv:1606.03179 [hep-ph]].
%%CITATION = doi:10.1140/epjc/s10052-017-4737-5;%%
%27 citations counted in INSPIRE as of 24 May 2019



%\cite{Maiani:2019cwl}
\bibitem{Maiani:2019cwl}
L.~Maiani, A.~D.~Polosa and V.~Riquer,
%``The Hydrogen Bond of QCD,''
arXiv:1903.10253 [hep-ph].
%%CITATION = ARXIV:1903.10253;%%
%1 citations counted in INSPIRE as of 24 May 2019



%\cite{Khachatryan:2016ydm}
\bibitem{Khachatryan:2016ydm}
V.~Khachatryan {\it et al.} [CMS Collaboration],
%``Observation of $\Upsilon$(1S) pair production in proton-proton collisions at $ \sqrt{s}=8 $ TeV,''
JHEP {\bf 1705}, 013 (2017) doi:10.1007/JHEP05(2017)013
[arXiv:1610.07095 [hep-ex]].
%%CITATION = doi:10.1007/JHEP05(2017)013;%%
%36 citations counted in INSPIRE as of 24 May 2019



\bibitem{S. Durgut}
S. Durgut (CMS), Search for Exotic Mesons at CMS (2018),
https://meetings.aps.org/Meeting/APR18/Session/ U09.6.


%\cite{Aaij:2018zrb}
\bibitem{Aaij:2018zrb}
R.~Aaij {\it et al.} [LHCb Collaboration],
%``Search for beautiful tetraquarks in the $\Upsilon(1S)\mu^+\mu^-$ invariant-mass spectrum,''
JHEP {\bf 1810}, 086 (2018) doi:10.1007/JHEP10(2018)086
[arXiv:1806.09707 [hep-ex]].
%%CITATION = doi:10.1007/JHEP10(2018)086;%%
%10 citations counted in INSPIRE as of 24 May 2019



%\cite{Iwasaki:1975pv}
\bibitem{Iwasaki:1975pv}
Y.~Iwasaki,
%``A Possible Model for New Resonances-Exotics and Hidden Charm,''
Prog.\ Theor.\ Phys.\ {\bf 54}, 492 (1975). doi:10.1143/PTP.54.492
%%CITATION = doi:10.1143/PTP.54.492;%%
%33 citations counted in INSPIRE as of 24 May 2019



%\cite{Chao:1980dv}
\bibitem{Chao:1980dv}
K.~T.~Chao,
%``The (c c) - (anti-c anti-c) (Diquark - anti-Diquark) States in e+ e- Annihilation,''
Z.\ Phys.\ C {\bf 7}, 317 (1981). doi:10.1007/BF01431564
%%CITATION = doi:10.1007/BF01431564;%%
%24 citations counted in INSPIRE as of 24 May 2019



%\cite{Ader:1981db}
\bibitem{Ader:1981db}
J.~P.~Ader, J.~M.~Richard and P.~Taxil,
%``Do Narrow Heavy Multi - Quark States Exist?,''
Phys.\ Rev.\ D {\bf 25}, 2370 (1982). doi:10.1103/PhysRevD.25.2370
%%CITATION = doi:10.1103/PhysRevD.25.2370;%%
%156 citations counted in INSPIRE as of 24 May 2019



%\cite{Zouzou:1986qh}
\bibitem{Zouzou:1986qh}
S.~Zouzou, B.~Silvestre-Brac, C.~Gignoux and J.~M.~Richard,
%``Four Quark Bound States,''
Z.\ Phys.\ C {\bf 30}, 457 (1986). doi:10.1007/BF01557611
%%CITATION = doi:10.1007/BF01557611;%%
%148 citations counted in INSPIRE as of 24 May 2019



%\cite{Heller:1986bt}
\bibitem{Heller:1986bt}
L.~Heller and J.~A.~Tjon,
%``On the Existence of Stable Dimesons,''
Phys.\ Rev.\ D {\bf 35}, 969 (1987). doi:10.1103/PhysRevD.35.969
%%CITATION = doi:10.1103/PhysRevD.35.969;%%
%59 citations counted in INSPIRE as of 24 May 2019



%\cite{SilvestreBrac:1992mv}
\bibitem{SilvestreBrac:1992mv}
B.~Silvestre-Brac,
%``Systematics of Q**2 (anti-Q**2) systems with a chromomagnetic interaction,''
Phys.\ Rev.\ D {\bf 46}, 2179 (1992). doi:10.1103/PhysRevD.46.2179
%%CITATION = doi:10.1103/PhysRevD.46.2179;%%
%38 citations counted in INSPIRE as of 24 May 2019



%\cite{SilvestreBrac:1993ry}
\bibitem{SilvestreBrac:1993ry}
B.~Silvestre-Brac and C.~Semay,
%``Spectrum and decay properties of diquonia,''
Z.\ Phys.\ C {\bf 59}, 457 (1993). doi:10.1007/BF01498626
%%CITATION = doi:10.1007/BF01498626;%%
%59 citations counted in INSPIRE as of 24 May 2019



%\cite{Bai:2016int}
\bibitem{Bai:2016int}
Y.~Bai, S.~Lu and J.~Osborne,
%``Beauty-full Tetraquarks,''
arXiv:1612.00012 [hep-ph].
%%CITATION = ARXIV:1612.00012;%%
%28 citations counted in INSPIRE as of 24 May 2019



%\cite{Anwar:2017toa}
\bibitem{Anwar:2017toa}
M.~N.~Anwar, J.~Ferretti, F.~K.~Guo, E.~Santopinto and B.~S.~Zou,
%``Spectroscopy and decays of the fully-heavy tetraquarks,''
Eur.\ Phys.\ J.\ C {\bf 78}, no. 8, 647 (2018)
doi:10.1140/epjc/s10052-018-6073-9 [arXiv:1710.02540 [hep-ph]].
%%CITATION = doi:10.1140/epjc/s10052-018-6073-9;%%
%23 citations counted in INSPIRE as of 24 May 2019



%\cite{Wang:2017jtz}
\bibitem{Wang:2017jtz}
Z.~G.~Wang,
%``Analysis of the $QQ\bar{Q}\bar{Q}$ tetraquark states with QCD sum rules,''
Eur.\ Phys.\ J.\ C {\bf 77}, no. 7, 432 (2017)
doi:10.1140/epjc/s10052-017-4997-0 [arXiv:1701.04285 [hep-ph]].
%%CITATION = doi:10.1140/epjc/s10052-017-4997-0;%%
%30 citations counted in INSPIRE as of 24 May 2019



%\cite{Wang:2018poa}
\bibitem{Wang:2018poa}
Z.~G.~Wang and Z.~Y.~Di,
%``Analysis of the vector and axialvector $QQ\bar{Q}\bar{Q}$ tetraquark states with QCD sum rules,''
arXiv:1807.08520 [hep-ph].
%%CITATION = ARXIV:1807.08520;%%
%1 citations counted in INSPIRE as of 24 May 2019



%\cite{Chen:2018cqz}
\bibitem{Chen:2018cqz}
W.~Chen, H.~X.~Chen, X.~Liu, T.~G.~Steele and S.~L.~Zhu,
%``Doubly hidden-charm/bottom $QQ\bar Q\bar Q$ tetraquark states,''
EPJ Web Conf.\ {\bf 182}, 02028 (2018)
doi:10.1051/epjconf/201818202028 [arXiv:1803.02522 [hep-ph]].
%%CITATION = doi:10.1051/epjconf/201818202028;%%
%3 citations counted in INSPIRE as of 24 May 2019



%\cite{Heupel:2012ua}
\bibitem{Heupel:2012ua}
W.~Heupel, G.~Eichmann and C.~S.~Fischer,
%``Tetraquark Bound States in a Bethe-Salpeter Approach,''
Phys.\ Lett.\ B {\bf 718}, 545 (2012)
doi:10.1016/j.physletb.2012.11.009 [arXiv:1206.5129 [hep-ph]].
%%CITATION = doi:10.1016/j.physletb.2012.11.009;%%
%60 citations counted in INSPIRE as of 24 May 2019



%\cite{Lloyd:2003yc}
\bibitem{Lloyd:2003yc}
R.~J.~Lloyd and J.~P.~Vary,
%``All charm tetraquarks,''
Phys.\ Rev.\ D {\bf 70}, 014009 (2004)
doi:10.1103/PhysRevD.70.014009 [hep-ph/0311179].
%%CITATION = doi:10.1103/PhysRevD.70.014009;%%
%31 citations counted in INSPIRE as of 24 May 2019



%\cite{Debastiani:2017msn}
\bibitem{Debastiani:2017msn}
V.~R.~Debastiani and F.~S.~Navarra,
%``A non-relativistic model for the $[cc][\bar{c}\bar{c}]$ tetraquark,''
Chin.\ Phys.\ C {\bf 43}, no. 1, 013105 (2019)
doi:10.1088/1674-1137/43/1/013105 [arXiv:1706.07553 [hep-ph]].
%%CITATION = doi:10.1088/1674-1137/43/1/013105;%%
%7 citations counted in INSPIRE as of 24 May 2019



%\cite{Barnea:2006sd}
\bibitem{Barnea:2006sd}
N.~Barnea, J.~Vijande and A.~Valcarce,
%``Four-quark spectroscopy within the hyperspherical formalism,''
Phys.\ Rev.\ D {\bf 73}, 054004 (2006)
doi:10.1103/PhysRevD.73.054004 [hep-ph/0604010].
%%CITATION = doi:10.1103/PhysRevD.73.054004;%%
%61 citations counted in INSPIRE as of 24 May 2019



%\cite{Berezhnoy:2011xy}
\bibitem{Berezhnoy:2011xy}
A.~V.~Berezhnoy, A.~K.~Likhoded, A.~V.~Luchinsky and
A.~A.~Novoselov,
%``Double J/psi-meson Production at LHC and 4c-tetraquark state,''
Phys.\ Rev.\ D {\bf 84}, 094023 (2011)
doi:10.1103/PhysRevD.84.094023 [arXiv:1101.5881 [hep-ph]].
%%CITATION = doi:10.1103/PhysRevD.84.094023;%%
%92 citations counted in INSPIRE as of 24 May 2019



%\cite{Berezhnoy:2011xn}
\bibitem{Berezhnoy:2011xn}
A.~V.~Berezhnoy, A.~V.~Luchinsky and A.~A.~Novoselov,
%``Tetraquarks Composed of 4 Heavy Quarks,''
Phys.\ Rev.\ D {\bf 86}, 034004 (2012)
doi:10.1103/PhysRevD.86.034004 [arXiv:1111.1867 [hep-ph]].
%%CITATION = doi:10.1103/PhysRevD.86.034004;%%
%29 citations counted in INSPIRE as of 24 May 2019



%\cite{Karliner:2016zzc}
\bibitem{Karliner:2016zzc}
M.~Karliner, S.~Nussinov and J.~L.~Rosner,
%``$Q Q \bar Q \bar Q$ states: masses, production, and decays,''
Phys.\ Rev.\ D {\bf 95}, no. 3, 034011 (2017)
doi:10.1103/PhysRevD.95.034011 [arXiv:1611.00348 [hep-ph]].
%%CITATION = doi:10.1103/PhysRevD.95.034011;%%
%44 citations counted in INSPIRE as of 24 May 2019



%\cite{Esposito:2018cwh}
\bibitem{Esposito:2018cwh}
A.~Esposito and A.~D.~Polosa,
%``A $bb\bar b\bar b$di-bottomonium at the LHC?,''
Eur.\ Phys.\ J.\ C {\bf 78}, no. 9, 782 (2018)
doi:10.1140/epjc/s10052-018-6269-z [arXiv:1807.06040 [hep-ph]].
%%CITATION = doi:10.1140/epjc/s10052-018-6269-z;%%
%10 citations counted in INSPIRE as of 24 May 2019



%\cite{Karliner:2017qhf}
\bibitem{Karliner:2017qhf}
M.~Karliner, J.~L.~Rosner and T.~Skwarnicki,
%``Multiquark States,''
Ann.\ Rev.\ Nucl.\ Part.\ Sci.\ {\bf 68}, 17 (2018)
doi:10.1146/annurev-nucl-101917-020902 [arXiv:1711.10626 [hep-ph]].
%%CITATION = doi:10.1146/annurev-nucl-101917-020902;%%
%44 citations counted in INSPIRE as of 24 May 2019



%\cite{Wu:2016vtq}
\bibitem{Wu:2016vtq}
J.~Wu, Y.~R.~Liu, K.~Chen, X.~Liu and S.~L.~Zhu,
%``Heavy-flavored tetraquark states with the $QQ\bar{Q}\bar{Q}$ configuration,''
Phys.\ Rev.\ D {\bf 97}, no. 9, 094015 (2018)
doi:10.1103/PhysRevD.97.094015 [arXiv:1605.01134 [hep-ph]].
%%CITATION = doi:10.1103/PhysRevD.97.094015;%%
%33 citations counted in INSPIRE as of 24 May 2019



%\cite{Chen:2019dvd}
\bibitem{Chen:2019dvd}
X.~Chen,
%``Analysis of hidden-bottom bb\bar{b}\bar{b} states,''
arXiv:1902.00008 [hep-ph].
%%CITATION = ARXIV:1902.00008;%%
%1 citations counted in INSPIRE as of 24 May 2019



%\cite{Liu:2019zuc}
\bibitem{Liu:2019zuc}
M.~S.~Liu, Q.~F.~LÃŒ, X.~H.~Zhong and Q.~Zhao,
%``Fully-heavy tetraquarks,''
arXiv:1901.02564 [hep-ph].
%%CITATION = ARXIV:1901.02564;%%
%2 citations counted in INSPIRE as of 24 May 2019



%\cite{Hughes:2017xie}
\bibitem{Hughes:2017xie}
C.~Hughes, E.~Eichten and C.~T.~H.~Davies,
%``Searching for beauty-fully bound tetraquarks using lattice nonrelativistic QCD,''
Phys.\ Rev.\ D {\bf 97}, no. 5, 054505 (2018)
doi:10.1103/PhysRevD.97.054505 [arXiv:1710.03236 [hep-lat]].
%%CITATION = doi:10.1103/PhysRevD.97.054505;%%
%19 citations counted in INSPIRE as of 24 May 2019



%\cite{Richard:2017vry}
\bibitem{Richard:2017vry}
J.~M.~Richard, A.~Valcarce and J.~Vijande,
%``String dynamics and metastability of all-heavy tetraquarks,''
Phys.\ Rev.\ D {\bf 95}, no. 5, 054019 (2017)
doi:10.1103/PhysRevD.95.054019 [arXiv:1703.00783 [hep-ph]].
%%CITATION = doi:10.1103/PhysRevD.95.054019;%%
%29 citations counted in INSPIRE as of 24 May 2019



%\cite{Czarnecki:2017vco}
\bibitem{Czarnecki:2017vco}
A.~Czarnecki, B.~Leng and M.~B.~Voloshin,
%``Stability of tetrons,''
Phys.\ Lett.\ B {\bf 778}, 233 (2018)
doi:10.1016/j.physletb.2018.01.034 [arXiv:1708.04594 [hep-ph]].
%%CITATION = doi:10.1016/j.physletb.2018.01.034;%%
%31 citations counted in INSPIRE as of 24 May 2019



%\cite{Wong:2001td}
\bibitem{Wong:2001td}
C.~Y.~Wong, E.~S.~Swanson and T.~Barnes,
%``Heavy quarkonium dissociation cross-sections in relativistic heavy ion collisions,''
Phys.\ Rev.\ C {\bf 65}, 014903 (2002) Erratum: [Phys.\ Rev.\ C {\bf
66}, 029901 (2002)] doi:10.1103/PhysRevC.66.029901,
10.1103/PhysRevC.65.014903 [nucl-th/0106067].
%%CITATION = doi:10.1103/PhysRevC.66.029901, 10.1103/PhysRevC.65.014903;%%
%100 citations counted in INSPIRE as of 24 May 2019



%\cite{SilvestreBrac:1996bg}
\bibitem{SilvestreBrac:1996bg}
B.~Silvestre-Brac,
%``Spectrum and static properties of heavy baryons,''
Few Body Syst.\ {\bf 20}, 1 (1996). doi:10.1007/s006010050028
%%CITATION = doi:10.1007/s006010050028;%%
%119 citations counted in INSPIRE as of 24 May 2019



%\cite{Tanabashi:2018oca}
\bibitem{Tanabashi:2018oca}
M.~Tanabashi {\it et al.} [Particle Data Group],
%``Review of Particle Physics,''
Phys.\ Rev.\ D {\bf 98}, no. 3, 030001 (2018).
doi:10.1103/PhysRevD.98.030001
%%CITATION = doi:10.1103/PhysRevD.98.030001;%%
%1686 citations counted in INSPIRE as of 24 May 2019

%\cite{Hiyama:2003cu}
\bibitem{Hiyama:2003cu}
E.~Hiyama, Y.~Kino and M.~Kamimura,
%``Gaussian expansion method for few-body systems,''
Prog.\ Part.\ Nucl.\ Phys.\ {\bf 51}, 223 (2003).
doi:10.1016/S0146-6410(03)90015-9
%%CITATION = doi:10.1016/S0146-6410(03)90015-9;%%
%338 citations counted in INSPIRE as of 27 May 2019


%\cite{Chen:2016jxd}
\bibitem{Chen:2016jxd}
W.~Chen, H.~X.~Chen, X.~Liu, T.~G.~Steele and S.~L.~Zhu,
%``Hunting for exotic doubly hidden-charm/bottom tetraquark states,''
Phys.\ Lett.\ B {\bf 773}, 247 (2017)
doi:10.1016/j.physletb.2017.08.034 [arXiv:1605.01647 [hep-ph]].
%%CITATION = doi:10.1016/j.physletb.2017.08.034;%%
%26 citations counted in INSPIRE as of 24 May 2019




\end{thebibliography}
\end{document}